\documentclass[twocolumn,aps,amsmath,prd,amssymb,eqsecnum,nofootinbib,superscriptaddress]{revtex4}

\DeclareMathOperator{\arctanh}{arctanh}

\newcommand{\bx}{\boldsymbol{x}}

\usepackage{graphicx}
\usepackage{multirow}
\usepackage{enumerate}
\usepackage{color}

\begin{document}

\title{Static Self-Forces in a Five-Dimensional Black Hole Spacetime}

\author{Peter Taylor}
\email{petertaylor@astro.cornell.edu}
\affiliation{Center for Radiophysics and Space Research, Cornell University, Ithaca, NY 14853, USA.}
\affiliation{School of Mathematical Sciences and Complex \& Adaptive Systems Laboratory, University College Dublin, UCD, Belfield, Dublin 4, Ireland.}
\author{\'Eanna \'E. Flanagan}
\email{flanagan@astro.cornell.edu}
\affiliation{Center for Radiophysics and Space Research, Cornell University, Ithaca, NY 14853, USA.}
\date{\today}
\begin{abstract}
We obtain the electric field and scalar field for a static point charge in closed form in the 5D Schwarzschild-Tangherlini black hole spacetime. We then compute the static self-force in each of these cases by assuming that the appropriate singular field is a 4D Hadamard Green's function on the constant time Riemannian slice. It is well known that the Hadamard representation of a Green's function involves an arbitrary regular biscalar $W_{0}(x,x')$, whose coincidence limit $w(x)$ appears in the expression for the self-force. We develop an axiomatic approach to reduce this arbitrary function to a single arbitrary dimensionless coefficient. We show that in the context of this approach to regularization, the self-force does not depend on any undetermined length-scale and need not depend on the internal structure of the charge. \end{abstract}
\maketitle

\section{Introduction}
The problem of motion in General Relativity is extremely complicated owing to the non-linearity of Einstein's equations. However, when there is a natural separation of scales, for example an extreme mass ratio binary black hole system, the problem is amenable to a perturbative analysis. To zeroth order in such a perturbative scheme, the small body is described by a point particle moving on a geodesic. Going beyond the geodesic approximation by including higher order corrections in the perturbative analysis can be interpreted as arising from the self-interaction of the point particle with its own field; this is the so-called self-force problem. In four dimensions, the self-force has enjoyed a long and fruitful history and is very well understood both at a formal level and at a practical computational level (see Ref.~\cite{PoissonLR} for a comprehensive review of the self-force problem and Ref.~\cite{WardellReview} for a review of computational approaches). Much of the attention in the literature is devoted to the development of a regularization prescription for curing the divergences that inevitably arise when working in the point-particle limit. Curing this kind of pathology is standard fare in quantum field theory where the infinities are absorbed into a renormalization of the constants which are then determined by observation. The difficulty in the classical theory is constructing the \textit{unique}, finite self-force that corresponds to the self-force on a finite-size body in the point-particle limit. There are a number of equivalent ways to regularize the self-force in curved spacetime dating back to the seminal paper of DeWitt and Brehme \cite{DeWittBrehme} who computed the formal expression for the regularized electromagnetic self-force, which was later corrected by Hobbs \cite{Hobbs}. Their approach relied heavily on the covariant decomposition of the Green's function for the electromagnetic vector potential into a ``direct'' and ``tail'' piece. The direct part, which they take to be the average of the advanced and retarded field, is singular and has support only on the light-cone while the tail part is regular with support only inside the light-cone. 

The most transparent derivation of the self-force is furnished by the method of matched asymptotic expansions which was utilized by Mino, Sasaki and Tanaka \cite{MiSaTa} to formally derive the self-force for a point mass in curved spacetime (see also Refs.~\cite{PoissonLR, DetweilerPRL}). A complementary derivation of the gravitational self-force was offered by Quinn and Wald \cite{QuinnWald} which was based on a simple set of physically motivated axioms. Hence, the equations of motion including first-order self-force effects are commonly referred to as the ``MiSaTaQuWa'' equations. In the same paper, Quinn and Wald also derive the electromagnetic self-force and their result agreed with that of DeWitt and Brehme \cite{DeWittBrehme} (taking into account the correction by Hobbs \cite{Hobbs}). This simple axiomatic approach was later extended to the scalar self-force by Quinn \cite{Quinn}. Each of the aforementioned calculations adopted the regularization scheme of DeWitt and Brehme \cite{DeWittBrehme}.  An alternative regularization prescription was offered by Detweiler and Whiting \cite{DetweilerWhiting2003} who instead decomposed the physical field into a ``singular'' and ``regular'' field and showed that the regular field is a solution to the homogeneous (source-free) wave equation and hence is smooth at the particle's location. This decomposition leads to a more natural interpretation of the self-force in terms of a particle's interaction with an external source-free field. Moreover, an axiomatic construction of the Detweiler-Whiting singular field was developed in Ref.~\cite{PoissonLR} by demanding that the singular Green's function is a symmetric, inhomogeneous solution to the wave equation with zero support inside the light-cone.

The gravitational self-force was given its most rigorous treatment by Gralla and Wald \cite{GrallaWald} who considered a one-parameter family of metrics describing an extended body that is scaled down to zero size and mass in an asymptotically self-similar manner. This derivation is essentially a more rigorous version
of the matched asymptotic derivation given by Mino, Sasaki and Tanaka \cite{MiSaTa}. A rigorous derivation of the electromagnetic self-force was later given by Gralla, Harte and Wald \cite{GrallaHarteWald}.

The self-force itself is not directly observable; instead the quantity of interest is the gravitational waveform produced by the emitted gravitational waves, for example by a particle inspiralling into a black hole. Computing the waveform itself requires solving the coupled problem of particle motion and radiation generation.  If one uses straightforward black hole perturbation theory, geodesic particle motion is associated with (i.e. arises at the same order as) first order metric perturbations. Accelerated particle motion and the first order self force are associated with second order metric perturbations.  This perturbation theory can self-consistently describe accelerated particle motion and the associated waveforms that are generated, but only over times short compared to the dephasing time (the timescale over which a geodesic orbit and the true orbit become out of phase by $\sim1$ cycle).  Thus, computing waveforms that are accurate over a radiation reaction timescale requires going beyond this kind of perturbation theory.  Two complementary approaches for achieving this are the self-consistent worldline approach of Pound \cite{Pound} and the two-timescale expansion technique of Hinderer and Flanagan \cite{HindererFlanagan}.

Despite the comprehensive work and progress in the self-force problem in four dimensions, the situation remains underdeveloped in higher dimensions. Obviously the intense endeavor over the last few decades has been driven by modeling extreme-mass-ratio inspirals and hence the self-force in a more abstract context has received little attention. Moreover, there is as yet no known general prescription or expression for self-forces in higher dimensions. Despite the lack of formal underpinning, there has been considerable recent interest in the self-force in higher dimensions \cite{Poisson5D, FrolovZelnikovSF, BirnholtzHadar1, BirnholtzHadar2} yielding some very unexpected results in odd dimensions. In particular, Beach, Poisson and Nickel \cite{Poisson5D} calculated the self-force on a static scalar and electric charge in a 5D black hole spacetime and found that the result depended on the radius of a sphere centered at the charge, which they interpret as the radius of the particle. Moreover, Frolov and Zelnikov \cite{FrolovZelnikovSF} calculated the self-force on a uniformly accelerating charge in Minkowski spacetime (or equivalently a static charge in Rindler spacetime) and their result depends on an undetermined infra-red cutoff length-scale which the authors postulate could be the scale over which the homogeneous gravitational field approximation is valid.

In this paper, we will revisit the calculation of a static electric and scalar charge in a 5D black hole spacetime. We derive closed-form representations of the electrostatic field and the static scalar field. In the absence of a rigorous derivation of the self-force in five dimensions, we attempt to use the Hadamard regularization prescription to compute a locally constructed singular field to subtract from the self-field. For a static charge in a static 5D spacetime, the field equations reduce to an elliptic wave equation on a 4D Riemannian manifold. It is well-known that the Hadamard form of the Green's function for this wave equation involves an arbitrary bi-scalar $W_{0}(x,x')$ that is undetermined by the local Hadamard expansion. For the physical field, this arbitrariness is a failure of a local expansion to encode global information such as boundary conditions or information about the quantum state in the quantum theory. For the singular field, we develop a set of axioms for constraining the coincidence limit of this bi-scalar, valid for static charges in an arbitrary static 5D spacetime. Remarkably, our axioms reduce a functions worth of arbitrariness down to a single arbitrary dimensionless coefficient. Using our closed-form expression for the physical field and our axiomatic construction of the singular field allows us to calculate simple closed-form expressions for the self-force in terms of this arbitrary coefficient. Our regularization scheme yields a self-force that does not depend on any undetermined length-scale such as an infra-red cutoff, unlike the calculation of Ref.~\cite{FrolovZelnikovSF}, nor does it depend on some particular model for the point particle, contrary to the result of Ref.~\cite{Poisson5D}.


\section{The Electrostatic Field}
\subsection{The Wave Equation}
The $(n+2)$-dimensional analogue of the Schwarzschild black hole, known as the Schwarzschild-Tangherlini metric \cite{Tangherlini}, is given by
\begin{align}
ds^{2}=-f(r)dt^{2}+\frac{1}{f(r)}dr^{2}+r^{2}d\Omega_{n}^{2},
\end{align}
where $f(r)=1-(r_{\textrm{\tiny{H}}}/r)^{n-1}$ with $r_{\textrm{\tiny{H}}}$ the horizon radius and $d\Omega_{n}^{2}$ is the line element on a unit $n$-sphere which may be defined inductively by
\begin{align}
d\Omega_{n}^{2}=\Omega_{A B}d x^{A}dx^{B}=d\theta_{1}+\sin^{2}\theta_{1}d\Omega_{n-1}^{2}.
\end{align}
The angular variables are $x^{A}=\{\theta_{1},...,\theta_{n-1},\phi\}$ with $\theta\in[0,\pi)$ and $\phi\in[0,2\pi)$. We can re-write the metric in terms of a lapse function and a Euclidean metric on a constant time hypersurface by
\begin{align}
ds^{2}=g_{ab}dx^{a}dx^{b}=-N^{2}dt^{2}+h_{\alpha\beta}dx^{\alpha}dx^{\beta},
\end{align}
where Latin indices $a, b=0,..,n+1$ are used for spacetime components and Greek indices $\alpha,\beta=1,...,n+1$ refer to tensor components on the constant time Riemannian slice. We are interested in computing the self-force on a static electric charge in this spacetime. The only non-trivial component of the vector potential is $\Phi:=\Phi_{t}$ and Maxwell's equations reduce to the (n+1)-dimensional Helmholtz equation on the Euclidean metric $h_{\alpha\beta}$
\begin{align}
\label{eq:fieldeqnelec}
\big\{\nabla^{2}-A^{\alpha}\partial_{\alpha}\big\}\Phi(\bx)=\Omega_{n}\,f\,j^{t}
\end{align}
where 
\begin{align}
\label{eq:vectorA}
A_{\alpha}=\frac{1}{2 f}\partial_{\alpha}f,\qquad \Omega_{n}=\frac{2\pi^{(n+1)/2}}{\Gamma(\tfrac{n+1}{2})},
\end{align}
and $\nabla^{2}$ is the Laplace operator on the metric $h_{\alpha\beta}$. The charge density for the point charge $e$ at position $\bx_{0}$ is given by the Dirac delta distribution
\begin{align}
j^{t}=\frac{e}{r_{0}^{n}}\delta(r-r_{0})\delta(\Omega,\Omega_{0}),
\end{align}
where $\delta(\Omega,\Omega_{0})=\delta^{n}(x^{A}-x_{0}^{A})/\sqrt{|\Omega_{AB}|}$ is the invariant Dirac delta distribution on an $n$-sphere. For a static particle in a static, spherically, symmetric spacetime the only non-vanishing component of the self-force is in the radial direction and is formally given by
\begin{align}
\label{eq:selfforceelecdef}
F^{r}=e \sqrt{f_{0}}\,\partial_{r} \Phi(\bx_{0}),
\end{align}
where $f_{0}=f(r_{0})$. This expression is meaningless as it stands since the field diverges at the position of the charge owing to the distributional nature of the point-particle source. Hence the gradient of the field must be regularized before evaluating at the position of the charge, which we discuss further in Sec.~\ref{sec:regularization}. 

Now, for the particular spacetime under consideration, the Green's function for the wave equation is
\begin{align}
\label{eq:waveeqnelec}
\big\{\nabla^{2}-A^{\alpha}\partial_{\alpha}\big\}G(\bx,\bx')=-\Omega_{n}\frac{\sqrt{f(r')}}{r'^{n}}\delta(r-r')\delta(\Omega,\Omega'),
\end{align}
which can be recast as
\begin{align}
\Big[r^{2}\frac{\partial^{2}}{\partial r^{2}}+n\,r\frac{\partial}{\partial r}+\frac{1}{f}D_{n}^{2}\Big]G(\bx,\bx')=-\Omega_{n}\frac{\delta(r-r')\delta(\Omega,\Omega')}{r'^{n-2}\sqrt{f(r')}}
\end{align}
with $D^{2}_{n}$ the Laplace operator on the $n$-sphere whose eigenfunctions are the generalized spherical harmonics \cite{Frye} $Y^{l,j}(\Omega)$ with eigenvalues $l(l+n-1)$. For a given $l$, there are
\begin{align}
\Lambda(l,n)=\frac{(2l+n-1)}{(n-1)!}\frac{(l+n-2)!}{l!}
\end{align}
linearly independent harmonics and the angular Dirac delta distribution may be decomposed in this basis as
\begin{align}
\delta(\Omega,\Omega')=\sum_{l=0}^{\infty}\sum_{j=0}^{\Lambda(l,n)-1}Y^{l,j}(\Omega)\bar{Y}^{l,j}(\Omega').
\end{align}
This suggests the following mode decomposition for the Green's function
\begin{align}
G(\bx,\bx')=\Omega_{n}\sum_{l=0}^{\infty}\sum_{j=0}^{\Lambda(l,n)-1}Y^{l,j}(\Omega)\bar{Y}^{l,j}(\Omega')g_{l}(r,r').
\end{align}
Since the spacetime is spherically symmetric, the radial Green's function $g_{l}(r,r')$ is independent of the mode number $j$ and hence we can apply the addition theorem
\begin{align}
\label{eq:legendreaddition}
\mathcal{P}_{l}(\cos\gamma_{n})=\frac{\Omega_{n}}{\Lambda(l,n)}\sum_{j=0}^{\Lambda(l,n)-1}Y^{l,j}(\Omega)\bar{Y}^{l,j}(\Omega'),
\end{align}
where $\gamma_{n}$ is the geodesic distance between two points on $\mathbb{S}^{n}$ and $\mathcal{P}_{l}$ is the generalized Legendre polynomial which is a solution to the differential equation
\begin{align}
\label{eq:legendreeqn}
\Big[\frac{d^{2}}{d\gamma_{n}^{2}}+\cot\gamma_{n}\frac{d}{d\gamma_{n}}+l(l+n-1)\Big]\mathcal{P}_{l}(\cos\gamma_{n})=0,
\end{align}
normalized so that
\begin{align}
\int_{0}^{\pi}\mathcal{P}_{l}(\cos\gamma_{n})\mathcal{P}_{l'}(\cos\gamma_{n})\,\sin^{n-1}\gamma_{n}\,d\gamma_{n}=\frac{\Omega_{n}}{\Omega_{n-1}}\frac{\delta_{l l'}}{\Lambda(l,n)}.
\end{align}
The electrostatic Green's function now assumes the form
\begin{align}
\label{eq:greensfnmode}
G(\bx,\bx')=\sum_{l=0}^{\infty}\Lambda(l,n)\mathcal{P}_{l}(\cos\gamma_{n})g_{l}(r,r'),
\end{align}
which upon substitution into the wave equation gives the following inhomogeneous ordinary differential equation satisfied by $g_{l}(r,r')$,
\begin{align}
\Big[r^{2}\frac{d^{2}}{dr^{2}}+n\,r\frac{d}{dr}-\frac{l(l+n-1)}{f}\Big]g_{l}(r,r')=-\frac{\delta(r-r')}{r'^{n-2}\sqrt{f(r')}}.
\end{align}
Rewriting this equation in terms of the dimensionless radius
\begin{align}
\label{eq:xi}
\xi=2(r/r_{\textrm{\tiny{H}}})^{n-1}-1,
\end{align}
leads to
\begin{align}
\Big[\frac{d}{d\xi}\Big((\xi+1)^{2}\frac{d}{d\xi}\Big)-\frac{l(l+n-1)}{(n-1)^{2}}\frac{\xi+1}{\xi-1}\Big]g_{l}(\xi,\xi')\nonumber\\
=-\frac{2}{(n-1)r_{\textrm{\tiny{H}}}^{n-1}}\frac{\delta(\xi-\xi')}{\sqrt{f(\xi')}}.
\end{align}
Following standard techniques (see \cite{MorseFeshbach1} for example), we can construct a solution to this inhomogeneous equation as a normalized product of homogeneous solutions,
\begin{align}
g_{l}(\xi,\xi')=\frac{2}{(n-1)r_{\textrm{\tiny{H}}}^{n-1}\sqrt{f(\xi')}}\frac{\Psi^{(1)}_{l}(\xi_{<})\Psi^{(2)}_{l}(\xi_{>})}{C_{l}},
\end{align}
where $\Psi^{(1)}_{l}(\xi)$ and $\Psi^{(2)}_{l}(\xi)$ are linearly independent solutions to the homogeneous equation satisfying regularity on the horizon and at infinity, respectively, and $C_{l}$ is determined by the Wronskian of these solutions. We have adopted the notation $\xi_{<}=\min\{\xi,\xi'\}$ and $\xi_{>}=\max\{\xi,\xi'\}$. The linearly independent solutions of the homogeneous equation are $\sqrt{f}\,P^{-1}_{l/(n-1)}(\xi)$ and $\sqrt{f}\,Q^{1}_{l/(n-1)}(\xi)$ where $P^{\mu}_{\nu}(z)$ and $Q^{\mu}_{\nu}(z)$ are the associated Legendre functions of the first and second kind, respectively. The former is the solution regular on the horizon while the latter is regular at infinity, i.e.,
\begin{align}
\Psi_{l}^{(1)}(\xi)&=\sqrt{f}\,P^{-1}_{\lambda}(\xi),\nonumber\\
\Psi_{l}^{(2)}(\xi)&=\sqrt{f}\,Q^{1}_{\lambda}(\xi),
\end{align}
where $\lambda=l/(n-1)$.

However, for $l=0$, these regularity conditions do not fix the solutions since both are everywhere regular and the choice is determined by enforcing the total charge as measured by an observer at infinity to be the charge of the point particle. This ambiguity is well known \cite{HanniRuffini, Linet:1976sq} and the appropriate choice of solutions is a constant for the inner solution and $1/(\xi+1)$ for the outer solution. We prefer to express these in terms of associated Legendre functions, i.e.,
\begin{align}
\Psi_{0}^{(1)}(\xi)&=\sqrt{f}P^{-1}_{0}(\xi)-2\sqrt{f}Q^{1}_{0}(\xi),\nonumber\\
\Psi_{0}^{(2)}(\xi)&=\sqrt{f}Q^{1}_{0}(\xi).
\end{align}
Noting that the normalization constant is simply $C_{l}=-1$ for all $l$, our mode-sum representation of the Green's function is
\begin{widetext}
\begin{align}
G(\bx,\bx')=-\frac{2\sqrt{f}}{(n-1)r_{\textrm{\tiny{H}}}^{n-1}}\sum_{l=0}^{\infty}\Lambda(l,n)\mathcal{P}_{l}(\cos\gamma_{n}) \,P^{-1}_{\lambda}(\xi_{<})Q^{1}_{\lambda}(\xi_{>})+\frac{4\sqrt{f}}{(n-1)r_{\textrm{\tiny{H}}}^{n-1}}\,Q^{1}_{0}(\xi)Q^{1}_{0}(\xi').
\end{align}
\end{widetext}
Finally, the convolution of the Green's function with the source term appearing in Eq.~(\ref{eq:fieldeqnelec}) yields
\begin{align}
\Phi(\bx)=-e\,\sqrt{f_{0}}\,G(\bx,\bx_{0}),
\end{align}
which results in the mode-sum representation of the electrostatic potential
\begin{align}
\Phi(\bx)=&\frac{2 e\,\sqrt{f}\sqrt{f_{0}}}{(n-1)r_{\textrm{\tiny{H}}}^{n-1}}\sum_{l=0}^{\infty}\Lambda(l,n)\mathcal{P}_{l}(\cos\gamma_{n})P^{-1}_{\lambda}(\xi_{<})Q^{1}_{\lambda}(\xi_{>})\nonumber\\
&-\frac{4\, e}{(n-1)r_{\textrm{\tiny{H}}}^{n-1}}\frac{1}{(\xi+1)(\xi_{0}+1)}
\end{align}
where we have used the fact that $f=(\xi-1)/(\xi+1)$ and $Q^{1}_{0}(\xi)=-1/\sqrt{\xi^{2}-1}$. In the four-dimensional Schwarzschild spacetime, we can sum these modes to retrieve the Copson-Linet \cite{Copson, Linet:1976sq} closed-form electrostatic field.

\subsection{Closed Form Static Field in Five Dimensions}
\label{sec:elecfieldclosed}
In this section, we specialize to the case of five dimensions ($n=3$). We will derive the closed-form representation of the electrostatic field by summing the mode-sum representation derived in the previous section. 

For $n=3$, the generalized Legendre function has the simple trigonometric representation
\begin{align}
\label{eq:legendre3}
\mathcal{P}_{l}(\cos\gamma_{3})=\frac{\sin(l+1)\gamma_{3}}{(l+1)\sin\gamma_{3}},
\end{align}
which is easily verified by substituting into Eq.~(\ref{eq:legendreeqn}). Henceforth, we shall drop the subscript on $\gamma_{3}$ for typographical convenience. The mode-sum representation of the field is now given by
\begin{align}
\label{eq:5Delecmodes}
\Phi(\bx)=&\frac{e\sqrt{f}\sqrt{f_{0}}}{r_{\textrm{\tiny{H}}}^{2}\sin\gamma}\sum_{l=0}^{\infty}(l+1)\,\sin(l+1)\gamma\,P^{-1}_{l/2}(\xi_{<})Q^{1}_{l/2}(\xi_{>})\nonumber\\
&-\frac{e\,r_{\textrm{\tiny{H}}}^{2}}{2r^{2}r_{0}^{2}}.
\end{align}
Now consider the standard Legendre addition formula \cite{GradRiz}
\begin{align}
Q_{\nu}(\xi\,\xi_{0}-(\xi^{2}-1)^{1/2}(\xi_{0}^{2}-1)^{1/2}\cos\Psi)\nonumber\\
=\sum_{k=-\infty}^{\infty}e^{-i k \pi}P^{-k}_{\nu}(\xi_{<})Q^{k}_{\nu}(\xi_{>})\cos k\Psi
\end{align}
valid for all $\nu\ne -1, -2, -3,...$, and taking the Fourier inverse allows us to express a product of associated Legendre functions in terms of a single Legendre function,
\begin{widetext}
\begin{align}
\label{eq:legendreproduct}
e^{-ik\pi}P^{-k}_{\nu}(\xi_{<})Q^{k}_{\nu}(\xi_{>})=\frac{1}{2\pi}\int_{0}^{2\pi}\cos k\Psi \,Q_{\nu}(\xi\,\xi_{0}-(\xi^{2}-1)^{1/2}(\xi_{0}^{2}-1)^{1/2}\cos\Psi)d\Psi.
\end{align}
Employing this identity in our expression for the field above yields
\begin{align}
\Phi(\bx)=&-\frac{e\,\sqrt{f}\sqrt{f_{0}}}{2\pi\,r_{\textrm{\tiny{H}}}^{2}\sin\gamma}\int_{0}^{2\pi}\cos\Psi\sum_{l=0}^{\infty}(l+1)\,\sin(l+1)\gamma\,Q_{l/2}(\xi\xi_{0}-(\xi^{2}-1)^{1/2}(\xi_{0}^{2}-1)^{1/2}\cos\Psi)d\Psi-\frac{e\,r_{\textrm{\tiny{H}}}^{2}}{2r^{2}r_{0}^{2}}.
\end{align}
In Ref.~\cite{Prudnikov3}, we find the summation formula
\begin{align}
\sum_{l=0}^{\infty}\cos(l+1)\gamma\,Q_{l/2}(z)=\frac{1}{(2z-2\cos 2\gamma)^{1/2}}\Big(\frac{\pi}{2}+\arctan \frac{2\cos \gamma}{(2 z-2\cos 2\gamma)^{1/2}}\Big).
\end{align}
Differentiating with respect to $\gamma$ gives
\begin{align}
\label{eq:sumq}
\sum_{l=0}^{\infty}(l+1)\,\sin(l+1)\gamma\,Q_{l/2}(z)=\frac{\sin\gamma}{(z-\cos 2\gamma)}+\frac{\pi \,\sin\gamma\cos\gamma}{\sqrt{2}(z-\cos 2\gamma)^{3/2}}+\frac{\sqrt{2}\,\sin\gamma\cos\gamma\arctan(\frac{2\cos\gamma}{\sqrt{2z-2\cos 2\gamma}})}{(z-\cos2\gamma)^{3/2}},
\end{align}
\end{widetext}
which permits us to sum the modes in our electrostatic potential to obtain the integral representation
\begin{align}
\label{eq:phielecint}
\Phi(\bx)&=-\frac{e\,\sqrt{f}\sqrt{f_{0}}}{2\pi\,r_{\textrm{\tiny{H}}}^{2}}\int_{0}^{2\pi}\Big(\frac{1}{\rho}
+\frac{\sqrt{2}\pi\,\cos\gamma}{\rho^{3/2}}\nonumber\\
&-\frac{\sqrt{2}\cos\gamma}{\rho^{3/2}}\arctan\Big(\frac{\sqrt{2\rho}}{2\cos\gamma}\Big)\Big)\cos\Psi\,d\Psi
-\frac{e\,r_{\textrm{\tiny{H}}}^{2}}{2r^{2}r_{0}^{2}},
\end{align}
where
\begin{align}
\label{eq:rho}
\rho=\xi\xi_{0}-\cos 2\gamma-(\xi^{2}-1)^{1/2}(\xi_{0}^{2}-1)^{1/2}\cos\Psi.
\end{align}
In arriving at (\ref{eq:phielecint}), we have restricted attention to $0\le\gamma\le\pi/2$ and made use of the identity,
\begin{align}
\label{eq:arctanid}
\arctan(x)=\tfrac{\pi}{2}-\arctan(1/x),\qquad x>0.
\end{align}
Defining
\begin{align}
\label{eq:chi}
\chi=\frac{\xi \xi_{0}-\cos 2\gamma}{(\xi^{2}-1)^{1/2}(\xi_{0}^{2}-1)^{1/2}},
\end{align}
we can recast our integrals as
\begin{align}
\Phi(\bx)=&-\frac{e\,\sqrt{f}\sqrt{f_{0}}}{2\pi\,r_{\textrm{\tiny{H}}}^{2}}\Bigg(\frac{1}{(\xi^{2}-1)^{1/2}(\xi_{0}^{2}-1)^{1/2}}\mathcal{I}^{(\textrm{e})}_{1}\nonumber\\
&+\frac{\sqrt{2}\pi\,\cos\gamma}{(\xi^{2}-1)^{3/4}(\xi_{0}^{2}-1)^{3/4}}\mathcal{I}^{(\textrm{e})}_{2}\nonumber\\
&-\frac{\sqrt{2}\,\cos\gamma}{(\xi^{2}-1)^{3/4}(\xi_{0}^{2}-1)^{3/4}}\mathcal{I}^{(\textrm{e})}_{3}\Bigg)-\frac{e\,r_{\textrm{\tiny{H}}}^{2}}{2r^{2}r_{0}^{2}},
\end{align}
where
\begin{align}
\mathcal{I}^{(\textrm{e})}_{1}&=\int_{0}^{2\pi}\frac{\cos\Psi}{(\chi-\cos\Psi)}d\Psi,\nonumber\\
\mathcal{I}^{(\textrm{e})}_{2}&=\int_{0}^{2\pi}\frac{\cos\Psi}{(\chi-\cos\Psi)^{3/2}}d\Psi,\nonumber\\
\mathcal{I}^{(\textrm{e})}_{3}&=\int_{0}^{2\pi}\frac{\arctan\Big[\frac{(\xi^{2}-1)^{1/4}(\xi_{0}^{2}-1)^{1/4}}{\sqrt{2}\,\cos\gamma}\sqrt{\chi-\cos\Psi}\Big]\cos\Psi}{(\chi-\cos\Psi)^{3/2}}d\Psi.
\end{align}
The first two of these are straight-forward to integrate; introducing the notation
\begin{align}
\label{eq:poles}
z_{\pm}=\chi\pm\sqrt{\chi^{2}-1},
\end{align}
then we have
\begin{align}
\mathcal{I}^{(\textrm{e})}_{1}&=2\pi\frac{z_{-}}{\sqrt{\chi^{2}-1}},\nonumber\\
\mathcal{I}^{(\textrm{e})}_{2}&=-2\frac{d}{d\chi}\int_{0}^{2\pi}\frac{\cos\Psi}{(\chi-\cos\Psi)^{1/2}}d\Psi\nonumber\\
&=-8\sqrt{2}\frac{d}{d\chi}\Big(\frac{K(z_{-})-E(z_{-})}{\sqrt{z_{-}}}\Big)\nonumber\\
&=\frac{4\sqrt{2}\sqrt{z_{+}}}{\sqrt{\chi^{2}-1}}\Big(-K(z_{-})+\frac{\chi}{\sqrt{\chi^{2}-1}}E(z_{-})\Big),
\end{align}
where $E(z)$, $K(z)$ are the complete Elliptic integral functions of the first and second kind, respectively. In arriving at the last line above, we made use of standard identities for derivatives of Elliptic integrals \cite{GradRiz} and we also used the fact that $z_{-}=1/z_{+}$.

The third of these integrals seems troublesome at first glance, but can be done by moving to the complex plane and treating the branch points with sufficient care. We make the following transformation
\begin{align}
z=e^{i\Psi}\quad\implies\quad d\Psi=\frac{dz}{i\,z}\,\,\,\textrm{and}\,\,\, \cos \Psi=\tfrac{1}{2}(z+1/z).
\end{align}
In terms of $z$, the integral goes around the unit circle in the complex plane and may be written as
\begin{align}
\label{eq:i3contour}
\mathcal{I}^{(\textrm{e})}_{3}=i\sqrt{2}\oint \frac{f(z)(z^{2}+1)}{ (z-z_{-})^{2}(z-z_{+})^{2}}dz,
\end{align}
where
\begin{widetext}
\begin{align}
\label{eq:fholomorphic}
f(z)= \sqrt{\frac{(z-z_{-})(z-z_{+})}{z}}\arctanh\Bigg(\frac{(\xi^{2}-1)^{1/4}(\xi_{0}^{2}-1)^{1/4}}{2\cos\gamma}\sqrt{\frac{(z-z_{-})(z-z_{+})}{z}}\Bigg),
\end{align}
\end{widetext}
and $z_{\pm}$ are defined in Eq.~(\ref{eq:poles}). One can show that $f(z)$ is holomorphic in a neighborhood of $z_{\pm}$ but has branch points at $z=0$ and at
\begin{align}
\label{eq:branchpoints}
\tilde{z}_{\pm}=\tilde{\chi}\pm\sqrt{\tilde{\chi}^{2}-1}
\end{align}
where
\begin{align}
\tilde{\chi}=\frac{\xi \xi_{0}+1}{(\xi^{2}-1)^{1/2}(\xi_{0}^{2}-1)^{1/2}}.
\end{align}
We note that $\tilde{z}_{-}<z_{-}$ and $\tilde{z}_{+}>z_{+}$ and hence introducing branch cuts on the real axis from $[0,\tilde{z}_{-}]$ and $[\tilde{z}_{+},\infty)$ yields a single-valued holomorphic integrand except at isolated simple poles at $z_{\pm}$. Of these poles, only $z_{-}$ lies inside the unit circle.

\begin{figure}
\centering
\includegraphics[width=9.0cm]{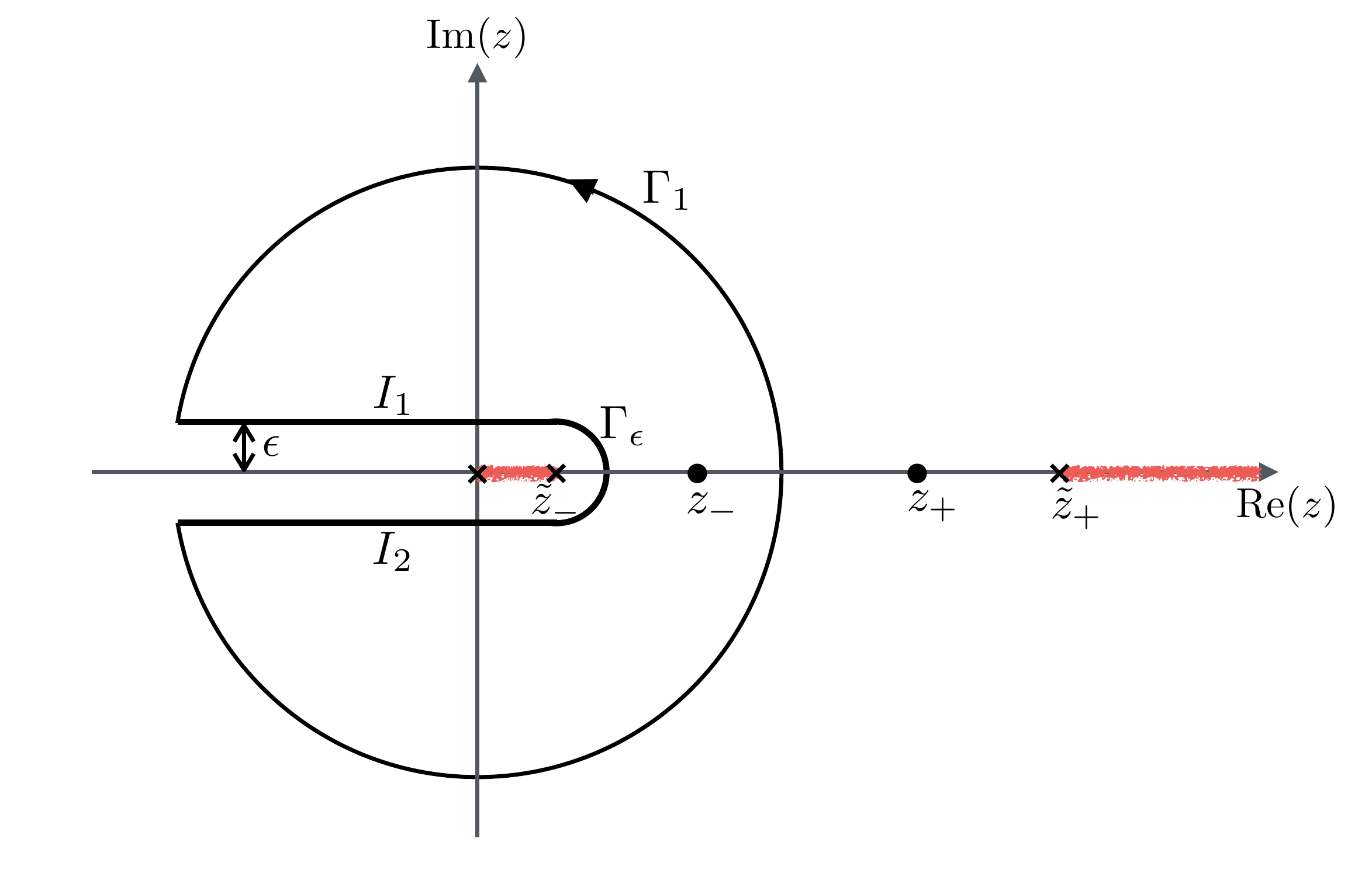}
\caption{{Plot of the branch cuts and the contour of integration for $\mathcal{I}_{3}^{(\textrm{e})}$.} }
\label{fig:contour}
\end{figure} 

Now we consider the integral around the deformed contour $\Gamma$ which consists of (see Fig.~\ref{fig:contour}):
\begin{enumerate}[(i)]
\item a circular arc $\Gamma_{1}$ traced counterclockwise from $-\sqrt{1-\epsilon^{2}}-i\epsilon$ to $-\sqrt{1-\epsilon^{2}}+i\epsilon$,
\item a horizontal line segment $I_{1}$ from $-\sqrt{1-\epsilon^{2}}+i\epsilon$ to $\tilde{z}_{-}+i\epsilon$,
\item a semi-circle $\Gamma_{\epsilon}$ of radius $\epsilon$ centered at $\tilde{z}_{-}$ and traced clockwise and
\item a horizontal line segment $I_{2}$ from $\tilde{z}_{-}-i\epsilon$ to $-\sqrt{1-\epsilon^{2}}-i\epsilon$.
\end{enumerate}
Hence the integral around the contour $\Gamma$ is given schematically by
\begin{align}
\int_{\Gamma}=\int_{\Gamma_{1}}+\int_{I_{1}}+\int_{\Gamma_{\epsilon}}+\int_{I_{2}}
\end{align}
which can be evaluated by the Cauchy Residue Theorem yielding
\begin{align}
\int_{\Gamma}=-\pi i \frac{(\xi^{2}-1)^{1/4}(\xi_{0}^{2}-1)^{1/4}}{\cos\gamma}\frac{\chi}{\sqrt{\chi^{2}-1}}.
\end{align}
In the limit as $\epsilon\to0$ the integral around $\Gamma_{1}$ tends to the closed contour around the unit circle that we required and so we obtain
\begin{align}
\label{eq:intcircle}
\oint=\lim_{\epsilon\to0}\int_{\Gamma_{1}}=-\pi i \frac{(\xi^{2}-1)^{1/4}(\xi_{0}^{2}-1)^{1/4}}{\cos\gamma}\frac{\chi}{\sqrt{\chi^{2}-1}}\nonumber\\
-\lim_{\epsilon\to0}\Big(\int_{I_{1}}+\int_{\Gamma_{\epsilon}}+\int_{I_{2}}\Big).
\end{align}
Following standard text-book techniques, it is straightforward to show that the integral around $\Gamma_{\epsilon}$ vanishes as $\epsilon\to 0$. Moreover, in this limit the line integrals $I_{1}$ and $I_{2}$ cancel in the left-half complex plane since the integrand is single-valued there, but we pick up a contribution from these line integrals across the cut since the $\arctanh$ is discontinuous across the cut and the difference above and below is $i\pi$. Therefore, we have
\begin{align}
\label{eq:intbranch}
\lim_{\epsilon\to 0}\Big(\int_{I_{1}}+\int_{I_{2}})&=\lim_{\epsilon\to 0}\Big(\int_{0+i\epsilon}^{\tilde{z}_{-}+i\epsilon}+\int_{\tilde{z}_{-}-i\epsilon}^{0-i\epsilon}\Big)\nonumber\\
&=-i\pi\int_{0}^{\tilde{z}_{-}}\frac{x^{2}+1}{\sqrt{x}(x^{2}-2 x \chi+1)^{3/2}} dx.
\end{align}
This integral can be written in terms of elliptic integrals as
\begin{align}
\label{eq:intbranchelliptic}
\int_{0}^{\tilde{z}_{-}}\frac{x^{2}+1}{\sqrt{x}(x^{2}-2 x \chi+1)^{3/2}} dx=-\frac{2\,\sqrt{z_{+}}}{(\chi^{2}-1)}\Big(\chi\,E(\psi, z_{-})\nonumber\\
-\sqrt{\chi^{2}-1}\,F(\psi, z_{-})\Big)\nonumber\\
+\frac{(\xi^{2}-1)^{1/4}(\xi_{0}^{2}-1)^{1/4}}{\cos\gamma}\frac{\chi(\chi-\tilde{z}_{-})}{\chi^{2}-1},
\end{align}
where
\begin{align}
\label{eq:psi}
\psi=\arcsin(\sqrt{\tilde{z}_{-}\,z_{+}}).
\end{align}
Combining Eqs.~(\ref{eq:intcircle})-(\ref{eq:intbranchelliptic}) with Eq.~(\ref{eq:i3contour}) yields
\begin{align}
\mathcal{I}^{(\textrm{e})}_{3}=\frac{\pi\sqrt{2}}{\chi^{2}-1}\Bigg[2\sqrt{z_{+}}\Big(\chi\,E\big(\psi, \,z_{-}\big)-\sqrt{\chi^{2}-1}\,F\big(\psi, \,z_{-}\big)\Big)\nonumber\\
+\frac{(\xi^{2}-1)^{1/4}(\xi_{0}^{2}-1)^{1/4}}{\cos\gamma}\chi\,(\tilde{z}_{-}-z_{-})\Bigg].
\end{align}
Finally, the electrostatic field is given by
\begin{align}
\label{eq:elecfieldclosedgamma}
&\Phi(\bx)=-\frac{e}{r_{\textrm{\tiny{H}}}^{2}(\xi+1)(\xi_{0}+1)\sqrt{\chi^{2}-1}}\Bigg[\frac{2 \chi^{2}-\chi\,\tilde{z}_{-}-1}{\sqrt{\chi^{2}-1}}\nonumber\\
&-\frac{2\,\sqrt{z_{+}}\,\cos\gamma}{(\xi^{2}-1)^{1/4}(\xi_{0}^{2}-1)^{1/4}}\Bigg\{2 \Bigg(K(z_{-})-\frac{\chi}{\sqrt{\chi^{2}-1}}E(z_{-})\Bigg)\nonumber\\
&-\Bigg(F\big(\psi, \,z_{-}\big)-\frac{\chi}{\sqrt{\chi^{2}-1}}E\big(\psi, \,z_{-}\big)\Bigg)\Bigg\}\Bigg]
\end{align}
where $\gamma$, $\xi$, $\chi$, $z_{\pm}$, $\tilde{z}_{\pm}$ and $\psi$ are defined in Eqs.~(\ref{eq:legendreaddition}), (\ref{eq:xi}), (\ref{eq:chi}), (\ref{eq:poles}), (\ref{eq:branchpoints}) and (\ref{eq:psi}), respectively. Now, recall that, since we have employed the identity (\ref{eq:arctanid}) to compute $\mathcal{I}_{3}^{(\textrm{e})}$, this expression is valid only for $0\le\gamma\le\pi/2$. We can analytically continue the Elliptic integral functions that appear in our expression for $\mathcal{I}_{3}^{(\textrm{e})}$ which results in the following representation for the field, valid over the entire range $0\le\gamma\le\pi$,
\begin{align}
\label{eq:elecfieldclosed}
\Phi(\bx)&=-\frac{e}{r_{\textrm{\tiny{H}}}^{2}(\xi+1)(\xi_{0}+1)\sqrt{\chi^{2}-1}}\Bigg[\frac{2 \chi^{2}-\chi\,\tilde{z}_{-}-1}{\sqrt{\chi^{2}-1}}\nonumber\\
&-\frac{2\,\sqrt{z_{+}}\,\cos\gamma}{(\xi^{2}-1)^{1/4}(\xi_{0}^{2}-1)^{1/4}}\Bigg\{2\,\Theta(\cos\gamma) \Bigg(K(z_{-})\nonumber\\
&-\frac{\chi}{\sqrt{\chi^{2}-1}}E(z_{-})\Bigg)-(2\,\Theta(\cos\gamma)-1)\Bigg(F\big(\psi, \,z_{-}\big)\nonumber\\
&-\frac{\chi}{\sqrt{\chi^{2}-1}}E\big(\psi, \,z_{-}\big)\Bigg)\Bigg\}\Bigg]
\end{align}
where $\Theta(z)$ is the Heaviside step function. We have checked that this field does indeed satisfy the wave equation and that it agrees numerically with its corresponding mode-sum representation.

\section{The Scalar Field}
In this section, we apply the previous analysis to a static scalar charge. The calculation is much the same as the electrostatic case and so we proceed with less details than before.

\subsection{The Wave Equation}
A static scalar particle with charge $q$ at position $\bx_{0}$ in a static, spherically symmetric $(n+2)$-dimensional spacetime satisfies the $(n+1)$-dimensional Helmholtz equation
\begin{align}
\label{eq:fieldeqnscalar}
\big\{\nabla^{2}+A^{\alpha}\partial_{\alpha}\big\}\varphi(\bx)=-\Omega_{n}\,\mu
\end{align}
where
\begin{align}
\mu=q\,\frac{\sqrt{f_{0}}}{r_{0}^{n}}\delta(r-r_{0})\delta(\Omega,\Omega_{0}).
\end{align}
The vector field $A^{\alpha}$ appearing in the potential is the same as in the electrostatic case (\ref{eq:vectorA}) but note that the potential has the opposite sign. The scalar self-force arising from the static scalar charge interacting with its own field has only a component in the radial direction and is formally given by
\begin{align}
\label{eq:selfforcescalardef}
F^{r}=q\,f_{0}\,\partial_{r}\varphi(\bx_{0}).
\end{align}
As before, we need to regularize the gradient of the field before evaluating at the location of the charge.
The corresponding Green's function for the static scalar wave equation is
\begin{align}
\label{eq:waveeqnscalar}
\big\{\nabla^{2}+A^{\alpha}\partial_{\alpha}\big\}G(\bx,\bx')\nonumber\\
=-\Omega_{n}\frac{\sqrt{f(r')}}{r'^{n}}\delta(r-r')\delta(\Omega,\Omega')
\end{align}
which is precisely Eq.~(\ref{eq:waveeqnelec}) with the transformation $A^{\alpha}\to-A^{\alpha}$. For the Schwarzschild-Tangherlini spacetime, this yields
\begin{align}
\Big\{r^{2}\frac{\partial^{2}}{\partial r^{2}}+r\Big(n+\frac{r\,f_{0}}{f}\Big)\frac{\partial}{\partial r}+\frac{1}{f}D_{n}^{2}\Big\}G(\bx,\bx')\nonumber\\
=-\frac{\Omega_{n}}{r'^{n-2}\sqrt{f(r')}}\delta(r-r')\delta(\Omega,\Omega').
\end{align}
The Green's function is again decomposed in a basis of generalized spherical harmonics as in Eq.~(\ref{eq:greensfnmode}), where now the radial part $g_{l}(r,r')$ satisfies
\begin{align}
\Big\{r^{2}\frac{d^{2}}{dr^{2}}+r\Big(n+\frac{r\,f_{0}}{f}\Big)\frac{d}{dr}-\frac{l(l+n-1)}{f}\Big\}g_{l}(r,r')\nonumber\\
=-\frac{\delta(r-r')}{r'^{n-2}\sqrt{f(r')}}.
\end{align}
In terms of the dimensionless radius $\xi$ defined in Eq.~(\ref{eq:xi}), this is
\begin{align}
\Big\{\frac{d}{d\xi}\Big( (\xi^{2}-1)\frac{d}{d\xi}\Big)-\frac{l(l+n-1)}{(n-1)^{2}}\Big\}g_{l}(\xi,\xi')\nonumber\\
=-\frac{2\sqrt{f(\xi')}}{(n-1)r_{\textrm{\tiny{H}}}^{n-1}}\delta(\xi-\xi').
\end{align}
The solutions of the corresponding homogeneous equation are Legendre functions of degree $\lambda=l/(n-1)$. We choose the Legendre function of the first kind $P_{\lambda}(\xi)$ to be the inner solution regular at the horizon and the Legendre function of the second kind $Q_{\lambda}(\xi)$ to be the outer solution regular at infinity. Unlike the electrostatic case, there is no monopole ambiguity and so the mode-sum representation of the Green's function is
\begin{align}
G(\bx,\bx')=\frac{2\sqrt{f(\xi')}}{(n-1)r_{\textrm{\tiny{H}}}^{n-1}}\sum_{l=0}^{\infty}\Lambda(l,n)\mathcal{P}_{l}(\cos\gamma_{n})P_{\lambda}(\xi_{<})Q_{\lambda}(\xi_{>}).
\end{align}
To obtain the field at $\bx$ due to a static charge at $\bx_{0}$, we convolve the Green's function with the scalar source which results in
\begin{align}
\varphi(\bx)=\frac{2\,q\,\sqrt{f_{0}}}{(n-1)r_{\textrm{\tiny{H}}}^{n-1}}\sum_{l=0}^{\infty}\Lambda(l,n)\mathcal{P}_{l}(\cos\gamma_{n})\,P_{\lambda}(\xi_{<})Q_{\lambda}(\xi_{>}).
\end{align}

\subsection{Closed Form Static Field in Five Dimensions}
For $n=3$, we adopt the representation (\ref{eq:legendre3}) for the generalized Legendre function which leads to the mode-sum representation of the static scalar field 
\begin{align}
\varphi(\bx)=\frac{q\,\sqrt{f_{0}}}{r_{\textrm{\tiny{H}}}^{2}\sin\gamma}\sum_{l=0}^{\infty}(l+1)\,\sin(l+1)\gamma\,P_{l/2}(\xi_{<})Q_{l/2}(\xi_{>}).
\end{align}
From Eq.~(\ref{eq:legendreproduct}), we have that
\begin{widetext}
\begin{align}
P_{l/2}(\xi_{<})Q_{l/2}(\xi_{>})=\frac{1}{2\pi}\int_{0}^{2\pi}Q_{l/2}(\xi\xi_{0}-(\xi^{2}-1)^{1/2}(\xi_{0}^{2}-1)^{1/2}\cos\Psi)\,d\Psi,
\end{align}
which allows us to recast the mode-sum as
\begin{align}
\varphi(\bx)=\frac{q\,\sqrt{f_{0}}}{2\pi\,r_{\textrm{\tiny{H}}}^{2}\sin\gamma}\int_{0}^{2\pi}\sum_{l=0}^{\infty}(l+1)\,\sin(l+1)\gamma\,Q_{l/2}(\xi\xi_{0}-(\xi^{2}-1)^{1/2}(\xi_{0}^{2}-1)^{1/2}\cos\Psi)\,d\Psi.
\end{align}
\end{widetext}
As in the electrostatic case, the sum can be performed using Eq.~(\ref{eq:sumq}) which results in the integral representation
\begin{align}
\varphi(\bx)=&\frac{q\,\sqrt{f_{0}}}{2\pi\,r_{\textrm{\tiny{H}}}^{2}}\int_{0}^{2\pi}\Bigg(\frac{1}{\rho}+\frac{\pi\sqrt{2}\,\cos\gamma}{\rho^{3/2}}\nonumber\\
&-\frac{\sqrt{2}\,\cos\gamma\arctan(\frac{\sqrt{2\rho}}{2\cos\gamma})}{\rho^{3/2}}\Bigg)d\Psi,
\end{align}
where $\rho$ is defined by Eq.~(\ref{eq:rho}). We write these integrals in terms of $\chi$ as
\begin{align}
\varphi(\bx)=\frac{q\,\sqrt{f_{0}}}{2\pi\,r_{\textrm{\tiny{H}}}^{2}}&\Bigg(\frac{1}{(\xi^{2}-1)^{1/2}(\xi_{0}^{2}-1)^{1/2}}\mathcal{I}^{(\textrm{s})}_{1}\nonumber\\
&+\frac{\pi\sqrt{2}\,\cos\gamma}{(\xi^{2}-1)^{3/4}(\xi_{0}^{2}-1)^{3/4}}\mathcal{I}^{(\textrm{s})}_{2}\nonumber\\
&-\frac{\sqrt{2}\,\cos\gamma}{(\xi^{2}-1)^{3/4}(\xi_{0}^{2}-1)^{3/4}}\mathcal{I}^{(\textrm{s})}_{3}\Bigg)
\end{align}
where
\begin{align}
\mathcal{I}^{(\textrm{s})}_{1}&=\int_{0}^{2\pi}\frac{1}{(\chi-\cos\Psi)}d\Psi,\nonumber\\
\mathcal{I}^{(\textrm{s})}_{2}&=\int_{0}^{2\pi}\frac{1}{(\chi-\cos\Psi)^{3/2}}d\Psi,\nonumber\\
\mathcal{I}^{(\textrm{s})}_{3}&=\int_{0}^{2\pi}\frac{\arctan[\frac{(\xi^{2}-1)^{1/4}(\xi_{0}^{2}-1)^{1/4}}{\sqrt{2}\,\cos\gamma}(\chi-\cos\Psi)^{1/2}]}{(\chi-\cos\Psi)^{3/2}}d\Psi.
\end{align}
The first integral can be performed in terms of elementary functions while the second results in a combination of complete Elliptic integral functions. Specifically, we have
\begin{align}
\mathcal{I}^{(\textrm{s})}_{1}&=\frac{2\pi}{\sqrt{\chi^{2}-1}}\nonumber\\
\mathcal{I}^{(\textrm{s})}_{2}&=-4\sqrt{2}\frac{\sqrt{z_{-}}}{\sqrt{\chi^{2}-1}}\Big(K(z_{-})-\frac{z_{+}}{\sqrt{\chi^{2}-1}}E(z_{-})\Big),
\end{align}
where $z_{\pm}$ are given by Eq.~(\ref{eq:poles}). The remaining integral $\mathcal{I}_{3}^{(\textrm{s})}$ is done by taking the integral around the unit circle in the complex plane
\begin{align}
\mathcal{I}_{3}^{(\textrm{s})}=i2\sqrt{2}\oint\frac{z\,f(z)}{(z-z_{+})^{2}(z-z_{-})^{2}}dz,
\end{align}
where $f(z)$ is defined in Eq.~(\ref{eq:fholomorphic}). We consider the same deformed contour as in the electrostatic case and again we pick up a contribution only from the pole at $z=z_{-}$ and a contribution across the branch cut which lies along $[0,\tilde{z}_{-}]$. The result is
\begin{align}
\mathcal{I}_{3}^{(\textrm{s})}
=&\frac{\pi\sqrt{2}}{\chi^{2}-1}\Bigg[2\sqrt{z_{-}}\Big(z_{+}\,E\big(\psi, \,z_{-}\big)-\sqrt{\chi^{2}-1}\,F\big(\psi, \,z_{-}\big)\Big)\nonumber\\
&+\frac{(\xi^{2}-1)^{1/4}(\xi_{0}^{2}-1)^{1/4}}{\cos\gamma}(\tilde{z}_{-}-z_{-})\Bigg],
\end{align}
where $\psi$ is given by Eq.~(\ref{eq:psi}).

Finally, the static scalar field is given by
\begin{align}
\label{eq:scalarfieldclosedgamma}
&\varphi(\bx)=\frac{q\sqrt{f_{0}}}{r_{\textrm{\tiny{H}}}^{2}(\xi^{2}-1)^{1/2}(\xi_{0}^{2}-1)^{1/2}\sqrt{\chi^{2}-1}}\Bigg[\frac{\chi-\tilde{z}_{-}}{\sqrt{\chi^{2}-1}}\nonumber\\
&-\frac{2\,\cos\gamma\,\sqrt{z_{-}}}{(\xi^{2}-1)^{1/4}(\xi_{0}^{2}-1)^{1/4}}\Bigg\{2\Big(K(z_{-})-\frac{z_{+}}{\sqrt{\chi^{2}-1}}E(z_{-})\Big)\nonumber\\
&-\Big(F\big(\psi, \,z_{-}\big)-\frac{z_{+}}{\sqrt{\chi^{2}-1}}E\big(\psi, \,z_{-}\big)\Big)\Bigg\}\Bigg].
\end{align}
where $\gamma$, $\xi$, $\chi$, $z_{\pm}$, $\tilde{z}_{\pm}$ and $\psi$ are defined in Eqs.~(\ref{eq:legendreaddition}), (\ref{eq:xi}), (\ref{eq:chi}), (\ref{eq:poles}), (\ref{eq:branchpoints}) and (\ref{eq:psi}), respectively. Again, this expression is valid only for $0\le\gamma\le\pi/2$ but can be analytically continued as before, resulting in
\begin{align}
\label{eq:scalarfieldclosed}
\varphi(\bx)&=\frac{q\sqrt{f_{0}}}{r_{\textrm{\tiny{H}}}^{2}(\xi^{2}-1)^{1/2}(\xi_{0}^{2}-1)^{1/2}\sqrt{\chi^{2}-1}}\Bigg[\frac{\chi-\tilde{z}_{-}}{\sqrt{\chi^{2}-1}}\nonumber\\
&-\frac{2\,\cos\gamma\,\sqrt{z_{-}}}{(\xi^{2}-1)^{1/4}(\xi_{0}^{2}-1)^{1/4}}\Bigg\{2\,\Theta(\cos\gamma)\Big(K(z_{-})\nonumber\\
&-\frac{z_{+}}{\sqrt{\chi^{2}-1}}E(z_{-})\Big)-(2\Theta(\cos\gamma)-1)\Big(F\big(\psi, \,z_{-}\big)\nonumber\\
&-\frac{z_{+}}{\sqrt{\chi^{2}-1}}E\big(\psi, \,z_{-}\big)\Big)\Bigg\}\Bigg].
\end{align}

We have checked that this field satisfies the static scalar wave equation and agrees numerically with its corresponding mode-sum representation.

\section{The Singular Field for Static Charges}
\subsection{Hadamard Green's Functions}
\label{sec:regularization}
The formal expression for the self-force on a point charge involves taking the gradient of a field evaluated at the location of the charge. This is clearly divergent as implied by the delta distribution source and therefore requires regularization. In four dimensions, the most elegant regularization prescription that results in the correct self-force is the Detweiler-Whiting \cite{DetweilerWhiting2003} construction of the singular field which yields, upon subtraction from the retarded field, a regular homogeneous solution to the wave equation. The Detweiler-Whiting Green's function is
\begin{align}
G_{\textrm{\tiny{DW}}}(x,x')=\frac{1}{2}\Big(\Delta^{1/2}(x,x')\,\delta(\sigma)-\Theta(\sigma)\,V(x,x')\Big),
\end{align}
where $\sigma(x,x')$ is Synge's \cite{Synge} world function which is half the square of the geodesic distance between $x$ and $x'$, $\Theta(z)$ is the step function, $\Delta(x,x')$ is the Van Vleck-Morette determinant and $V(x,x')$ is a regular, symmetric bi-solution of the homogeneous equation. In even dimensions, it is a straight-forward matter to construct the higher dimensional analogue of the Detweiler-Whiting Green's function, namely,
\begin{align}
G_{\textrm{\tiny{DW}}}(x,x')=\frac{\pi}{2}\frac{\Omega_{d-2}}{(2\pi)^{d/2}}\Big((-1)^{d/2-2}\delta^{(d/2-2)}(\sigma)\,U(x,x')\nonumber\\
-\frac{1}{\Gamma(\tfrac{d}{2}-1)}\Theta(\sigma)\,V(x,x')\Big).
\end{align}
This Green's function can be constructed axiomatically \cite{PoissonLR} by demanding that the appropriate parametrix satisfy
\begin{enumerate}
\item $(\Box-\xi\,R)G_{\textrm{\tiny{S}}}(x,x')=-\Omega_{d-2}\delta(x,x')$,
\item $G_{\textrm{\tiny{S}}}(x,x')=G_{\textrm{\tiny{S}}}(x',x)$,
\item $G_{\textrm{\tiny{S}}}(x,x')=0$,\qquad \textrm{if}\qquad $\sigma(x,x')<0$.
\end{enumerate}
In odd dimensions, by contrast, it seems likely that no Green's function satisfying these properties exists, even in Minkowski spacetime; a variety of simple choices of ansatz can be shown to not satisfy the properties above.

If we are only interested in a static charge in a static spacetime, then the wave equation reduces to an elliptic equation on a Riemannian manifold, which has a unique solution subject to boundary conditions. Although the Green's function for the physical field is sensitive to global properties such as boundary conditions, within some local neighborhood all Green's functions for all choices of boundary conditions are described by the same universal Hadamard form, which we define below. Each propagator within this family of Hadamard Green's functions is parametrized by a particular choice for some regular biscalar $W_{0}(x,x')$. It is the difference between this biscalar $W_{0}(x,x')$ for the physical field and the singular field that gives rise to the self-force.  We assume that our singular field constructed from a Hadamard Green's function (for some judicious choice of $W_{0}(x,x')$) does not exert a force on the charge and hence subtracting from the physical field and taking the gradient leads to the correct self-force. This assumption can be partially justified on the basis of Harte's formalism \cite{HarteReview, Harte2010, Harte2009, Harte2008} for self-fields on extended bodies wherein it is proved that the self-force contribution from a singular field obtained from a geometrically constructed symmetric two-point function acts only to renormalize the moments of the body. Hence our results for the self-force ought to be correct up to such renormalizations.

In his seminal lectures, Hadamard \cite{Hadamard} constructed a local solution to an arbitrary second order linear partial differential equation by assuming the solution has a series expansion in Synge's world function $\sigma(x,x')$. The expansion is only valid in a neighborhood where $x$ and $x'$ are connected by a unique geodesic, the so-called normal convex neighborhood. We present only a brief overview here and we refer the reader to the detailed description of Hadamard renormalization in arbitrary dimensions given in Ref.~\cite{Decanini:2008}. 

In $d$ odd dimensions, the universal Hadamard form for a Green's function takes the form
\begin{align}
\label{eq:hadamardodd}
G_{\textrm{\tiny{H}}}(x,x')\sim \frac{U(x,x')}{\sigma^{d/2-1}}+W(x,x'),
\end{align}
where $U(x,x')$ and $W(x,x')$ are regular biscalars and $U$ satisfies coincidence boundary condition $[U]=1$ (we have adopted square brackets around a bi-tensor to denote the coincidence limit $x\to x'$). Substituting the \textit{ansatz}
\begin{align}
U(x,x')=\sum_{k=0}^{\infty}U_{k}(x,x')\sigma^{k}\nonumber\\
W(x,x')=\sum_{k=0}^{\infty}W_{k}(x,x')\sigma^{k}
\end{align}
into the wave equation yields a set of recursive transport equations for the coefficients $U_{k}$ and $W_{k}$. The recursion relations completely determine all of the $U_{k}$ coefficients and also determine all of the $W_{k}$ coefficients except for $W_{0}$ which remains arbitrary. This implies that the Hadamard representation of a Green's function is not completely determined by a local expansion. For the physical field, $W_{0}$ encodes global information such as boundary conditions. In quantum field theory, for the Hadamard representation of the two-point function, $W_{0}$ encodes the information about the quantum state of the system.

In $d$ even dimensions, the Hadamard form is given by
\begin{align}
\label{eq:hadamardeven}
G_{\textrm{\tiny{H}}}(x,x')\sim\frac{U(x,x')}{\sigma^{d/2-1}}+V(x,x')\,\log\sigma+W(x,x')
\end{align}
where $U(x,x')$, $V(x,x')$ and $W(x,x')$ are regular biscalars. One can make the $\log$ term explicitly dimensionless by introducing a length scale $s$, and writing it as $V \log \sigma/s^{2}$ but this length scale can be absorbed into a redefinition of $W(x,x')$. The biscalar $U(x,x')$ again satisfies the boundary condition $[U]=1$, but unlike the odd-dimensional case now possesses a finite series expansion
\begin{align}
U(x,x')=\sum_{k=0}^{d/2-2}U_{k}(x,x')\sigma^{k}.
\end{align}
Assuming the series expansions
\begin{align}
\label{eq:VWseries}
V(x,x')&=\sum_{k=0}^{\infty}V_{k}(x,x')\sigma^{k},\nonumber\\
W(x,x')&=\sum_{k=0}^{\infty}W_{k}(x,x')\sigma^{k},
\end{align}
and substituting into the wave equation determines the $U_{k}$ and $V_{k}$ completely. As before, this determines all of the $W_{k}$ coefficients except for $W_{0}$.

If the wave equation is self-adjoint, say,
\begin{align}
\Big\{\nabla^{2}-P(x)\Big\}G_{\textrm{\tiny{H}}}(x,x')=-\Omega_{d-2}\,\delta(x,x')
\end{align}
where $P(x)$ is a potential that does not contain any derivative operators, then $U(x,x')$, $V(x,x')$ and $W(x,x')$ are symmetric biscalars. The biscalar $W(x,x')$ is then a bi-solution of the homogeneous wave equation in odd dimensions while $V(x,x')$ is a bi-solution of the homogeneous wave equation in even dimensions. Furthermore, we have that (see, for example, \cite{Christensen:1976vb, BrownOttewill1986})
\begin{align}
\label{eq:vanvleckexp}
U_{0}(x,x')=\Delta^{1/2}(x,x')=1+\tfrac{1}{12}R_{\alpha'\beta'}\sigma^{;\alpha'}\sigma^{;\beta'}\nonumber\\
-\tfrac{1}{24}R_{\alpha'\beta';\gamma'}\sigma^{;\alpha'}\sigma^{;\beta'}\sigma^{;\gamma'}+\textrm{O}(\sigma^{2})
\end{align}
while $V_{0}$ and $W_{0}$ possess covariant Taylor expansions of the form
\begin{align}
\label{eq:vwexp}
V_{0}(x,x')&=v(x')-\tfrac{1}{2}v_{;a'}(x')\sigma^{;a'}+\textrm{O}(\sigma),\nonumber\\
W_{0}(x,x')&=w(x')-\tfrac{1}{2}w_{;a'}(x')\sigma^{;a'}+\textrm{O}(\sigma),
\end{align}
where $v(x')=[V(x,x')]$ and $w(x')=[W(x,x')]$. Explicit expressions for $v(x')$ are obtained by taking coincidence limits of its transport equation, for example, for $d=4$ we have $v(x')=\tfrac{1}{2}(P(x')-\tfrac{1}{6} R(x'))$.

\subsection{Constraining $w(x)$ for a Static Charge in a 5D Static Spacetime}
For a static charge located at $\bx_{0}$ in a 4D static spacetime, it has been shown \cite{CasalsPoissonVega} that the Detweiler-Whiting Green's function is equivalent to the three-dimensional Hadamard Green's function on a constant time slice with $w(\bx_{0})=0$, at least up to the order required for the computation of the self-force (assuming the singular field is derived from a symmetric Hadamard Green's function, then only $w(\bx_{0})$ and its derivative appear in the self-force). For ultrastatic spacetimes ($g_{tt}=-1$), they are equivalent up to all orders with $W(\bx,\bx_{0})=0$. In these cases, it is possible to determine $w(\bx)$ by direct comparison with the known Detweiler-Whiting field. In this section, we address whether we can determine $w(\bx)$ without knowing the corresponding Detweiler-Whiting field. We note that in Ref.~\cite{Poisson5D}, the authors adopt a definition for the self-force that involves a spherical averaging procedure and the smooth part of the Green's function does not contribute to the self-force after this averaging is performed. Thus, their analysis effectively corresponds to choosing the singular field for which $W_0(\bx, \bx_0)=0$. However, such a spherical averaging results in a self-force that depends on the details of the model charge. For example, one can model the point particle as a sphere of constant proper radius centered at the charge's location or alternatively as a sphere of constant proper volume centered at the charge's location; the resultant self-force is different for each model. Moreover, the self-force diverges as the radius of this sphere shrinks to zero. 

Consider now a static electric charge in a $d$-dimensional static spacetime. We decompose the metric into a lapse function $N$ and a $(d-1)$-dimensional Riemannian metric $h_{\alpha\beta}$. The only non-trivial component of the vector potential is the time component $\Phi_{t}=\Phi$ which satisfies
\begin{align}
\big\{\nabla^{2}-A^{\alpha}\partial_{\alpha}\big\}\Phi(\bx)=\Omega_{d-2}\,N^{2}\,j^{t}
\end{align}
where $\nabla^{2}=h^{\alpha\beta}\nabla_{\alpha}\nabla_{\beta}$ and
\begin{align}
\label{eq:staticsource}
A_{\alpha}=\frac{1}{N}\partial_{\alpha}N,\qquad j^{t}=e\,N_{0}^{-1}\,\delta_{d-1}(\bx,\bx_{0}).
\end{align}
We wish to rewrite this wave-equation in self-adjoint form. This ensures that the corresponding Green's function will be symmetric in its arguments and hence our singular field will exert no force on the charge, but acts only to renormalize the moments of the body \cite{HarteReview}. We can express the electrostatic wave equation as
\begin{align}
\label{eq:waveeqndalembertian}
\frac{1}{N\sqrt{h}}\frac{\partial}{\partial x^{\alpha}}\Big(\frac{\sqrt{h}\,h^{\alpha\beta}}{N}\frac{\partial\Phi}{\partial x^{\beta}}\Big)=\Omega_{d-2}\,j^{t},
\end{align}
which suggests that electrostatic potential satisfies a Poisson equation on a conformally related metric with a rescaled charge. To see this, we choose our conformal metric $\tilde{h}_{\alpha\beta}$ to satisfy
\begin{align}
\sqrt{\tilde{h}}\,\tilde{h}^{\alpha\beta}=\frac{\sqrt{h}\,h^{\alpha\beta}}{N},
\end{align}
which in turn implies
\begin{align}
\label{eq:conformalmetricelectric}
\tilde{h}_{\alpha\beta}=N^{-2/(d-3)}h_{\alpha\beta},
\end{align}
whence Eq.~(\ref{eq:waveeqndalembertian}) may be recast into the Poisson form
\begin{align}
\label{eq:poisson5D}
\tilde{\nabla}^{2}\Phi=e\,\Omega_{d-2}\tilde{\delta}_{d-1}(\bx,\bx_{0}),
\end{align}
where $\tilde{\nabla}^{2}$ is the D'Alembertian operator with respect to the conformal metric $\tilde{h}_{\alpha\beta}$ and $\tilde{\delta}(\bx, \bx_{0})=\delta(\bx-\bx_{0})/\sqrt{\tilde{h}}$.

Restricting attention to a static charge in a 5D spacetime, then $\tilde{h}_{\alpha\beta}=N^{-1}h_{\alpha\beta}$ and the Hadamard Green's function corresponding to the operator $\tilde{\nabla}^{2}$ has the form 
\begin{align}
\label{eq:Gelec}
\tilde{G}_{\textrm{\tiny{H}}}(\bx, \bx')=\frac{1}{4}\Big(\frac{\tilde{\Delta}^{1/2}(\bx, \bx')}{\tilde{\sigma}}+\tilde{V}(\bx, \bx')\,\log\tilde{\sigma}+\tilde{W}(\bx,\bx')\Big)
\end{align}
where all biscalars appearing here are with respect to $\tilde{h}_{\alpha\beta}$. In light of the expansions (\ref{eq:vanvleckexp})-(\ref{eq:vwexp}), we have that
\begin{align}
\label{eq:vwpoissonexp}
\tilde{\Delta}^{1/2}(\bx,\bx')&=1+\tfrac{1}{12}\tilde{R}_{\alpha'\beta'}\tilde{\sigma}^{;\alpha'}\tilde{\sigma}^{;\beta'}-\tfrac{1}{24}\tilde{R}_{\alpha'\beta';\gamma'}\tilde{\sigma}^{;\alpha'}\tilde{\sigma}^{;\beta'}\tilde{\sigma}^{;\gamma'}\nonumber\\
&\quad+\textrm{O}(\tilde{\sigma}^{2})\nonumber\\
\tilde{V}_{0}(\bx,\bx')&=-\tfrac{1}{12}\tilde{R}(\bx')+\tfrac{1}{24}\tilde{R}_{;a'}(\bx')\tilde{\sigma}^{;a'}+\textrm{O}(\tilde{\sigma}),\nonumber\\
\tilde{W}_{0}(\bx,\bx')&=\tilde{w}(\bx')-\tfrac{1}{2}\tilde{w}_{;a'}(\bx')\tilde{\sigma}^{;a'}+\textrm{O}(\tilde{\sigma}).
\end{align}
The singular field can be obtained by a trivial convolution of this conformal Green's function $\tilde{G}_{\textrm{\tiny{H}}}$ against the source term in (\ref{eq:poisson5D}), 
\begin{align}
\label{eq:singfieldelec}
\Phi_{\textrm{\tiny{S}}}(\bx)&=-e\,\tilde{G}_{\textrm{\tiny{H}}}(\bx,\bx_{0})\nonumber\\
&=-\frac{e}{4}\Big(\frac{\tilde{\Delta}^{1/2}(\bx, \bx_{0})}{\tilde{\sigma}}+\tilde{V}(\bx, \bx_{0})\,\log\tilde{\sigma}+\tilde{W}(\bx,\bx_{0})\Big).
\end{align}

Now computing the self-force requires evaluating the gradient of the regular field $\Phi_{\textrm{R}}=\Phi-\Phi_{\textrm{\tiny{S}}}$ before taking the coincident limit $\bx\to\bx_{0}$. Hence only $\tilde{w}(\bx_{0})=\tilde{w}_{0}$ and its derivative appear in the expression for the self-force. Notwithstanding the arbitrariness of this function which is necessary to encode different boundary conditions in the physical field, one still expects the singular field to be insensitive to such boundary conditions and hence $\tilde{w}(\bx)$ for the singular Green's function ought to be locally constructed. We postulate the following axioms on $\tilde{w}$:
\begin{enumerate}
\item \textbf{$\tilde{w}$ must transform appropriately under length rescalings}. Explicitly, under a change of units of length by a factor $l$, we have that $\tilde{\sigma}\to l^{2}\tilde{\sigma}$, $\tilde{V}\to l^{-2}\tilde{V}$ and $\tilde{G}_{\textrm{\tiny{H}}}\to l^{-2}\tilde{G}_{\textrm{\tiny{H}}}$ and hence by Eq.~(\ref{eq:Gelec})
\begin{align}
\tilde{w}\to l^{-2}(\tilde{w}-\tilde{v}\log l^{2}).
\end{align}
This implies that $\tilde{w}$ assumes the form
\begin{align}
\tilde{w}=\tilde{v}\,\log |Q_{1}|+Q_{2},
\end{align}
where $Q_{1}$ and $Q_{2}$ are scalars with dimensions inverse length squared $[L]^{-2}$.
\item \textbf{$\tilde{w}$ must be constructed only from local geometrical quantities}. This is completely reasonable since the singular field should only be sensitive to the local physics near the particle. Since the singular field satisfies the Poisson equation (\ref{eq:poisson5D}), the only geometric scalars in the theory are curvature invariants of $\tilde{h}_{\alpha\beta}$. Since, the only curvature invariant with dimensions $[L]^{-2}$ is the Ricci scalar on $\tilde{h}_{\alpha\beta}$, the possible terms appearing in $Q_{1}$ and $Q_{2}$ are of the form $F(K)\,\tilde{R}$, where $K$ is a dimensionless ratio of curvature scalars, e.g., $\tilde{R}_{ab}\tilde{R}^{ab}/\tilde{R}^{2}$, and $F$ is an arbitrary dimensionless function of such ratios.

\item \textbf{Finally, we assume that $\tilde{w}$ does not depend on ratios of curvature scalars}. Unlike the previous two axioms, this is a strong assumption.  Obviously we want our prescription to be valid for arbitrary static spacetimes and the most straight-forward way to achieve this is to rule out such curvature ratios. The more conservative way to achieve this would be to impose analyticity in $K$ in some complex neighborhood of the positive real axis (see, for example, \cite{WaldQFT} for a discussion in the context of quantum field theory). It turns out that this restriction is not strong enough to give any predictive power since there is still an infinite space of functions that satisfy this analyticity criterion \cite{TichyFlanagan}, for example, $F(K)=K^{2}/(1+K^{4})$ and $F(K)=K^{2}\exp(-K^{2})$ are smooth functions satisfying $F(K)\to0$ as $K\to0$ and as $K\to\infty$. This axiom implies that
\begin{align}
Q_{1,2}=\alpha_{1,2}\,\tilde{R},
\end{align}
where $\alpha_{1,2}$ are arbitrary dimensionless constants.
\end{enumerate}
Now consideration of the above axioms necessitates that
\begin{align}
\tilde{w}=\tilde{v}\,\log \alpha_{1}\,|\tilde{R}|+\alpha_{2}\,\tilde{R},\qquad \alpha_{1}>0.
\end{align}
There is still some redundancy in this construction. We note from Eq.~(\ref{eq:vwpoissonexp}) that $\tilde{v}(\bx)=\tilde{V}_{0}(\bx,\bx)=-\tfrac{1}{12}\tilde{R}(\bx)$, and hence we can redefine our constants such that 
\begin{align}
\tilde{w}=\tilde{v}\,\log\,\alpha_{1}|\tilde{v}|+\alpha_{2}\,\tilde{v},\qquad\alpha_{1}>0.
\end{align}
Finally, we can absorb $\alpha_{1}$ into a redefinition of $\alpha_{2}$,  for example, defining $\alpha_{2}=\alpha-\log\alpha_{1}$ results in
\begin{align}
\label{eq:wreduced}
\tilde{w}=\tilde{v}\,\log \,|\tilde{v}|+\alpha\,\tilde{v}.
\end{align}
The crux of this analysis is that we have reduced the arbitrariness contained in the function $\tilde{w}(\bx)$ down to a single arbitrary dimensionless constant for an arbitrary static 5D spacetime. The constant does not depend on the particular spacetime.

The previous analysis applies almost \textit{verbatim} for a minimally coupled scalar field (or a non-minimally coupled field in a vacuum spacetime) so we omit much of the details. The field equation for a static scalar charge in a static 5D spacetime can be recast into a Poisson equation
\begin{align}
\tilde{\nabla}^{2}\varphi(\bx)=q\,N_{0}\,\tilde{\delta}(\bx,\bx_{0}),
\end{align}
where now the wave operator is with respect to the metric
\begin{align}
\label{eq:conformalmetricscalar}
\tilde{h}_{\alpha\beta}=N\,h_{\alpha\beta}.
\end{align}
The singular field has the form
\begin{align}
\label{eq:singfieldscalar}
\varphi_{\textrm{\tiny{S}}}(\bx)=\frac{q}{4}\, N_{0}\,\Big(\frac{\tilde{\Delta}^{1/2}(\bx,\bx_{0})}{\tilde{\sigma}}+\tilde{V}(\bx, \bx_{0})\,\log\tilde{\sigma}\nonumber\\
+\tilde{W}(\bx,\bx_{0})\Big),
\end{align}
where all quantities are functions of the metric (\ref{eq:conformalmetricscalar}). The axioms we applied to the electrostatic case in order to reduce the arbitrariness of the coincidence limit of $\tilde{W}_{0}(\bx,\bx')$ apply also to the scalar case. Hence, $\tilde{w}(\bx)$ has precisely the same form as in the electrostatic case (\ref{eq:wreduced}) but with
\begin{align}
\label{eq:v0scalar}
\tilde{v}=-\tfrac{1}{12}\,\tilde{R},
\end{align}
where $\tilde{R}$ is now the Ricci scalar of the metric (\ref{eq:conformalmetricscalar}).

\subsection{Singular Field for a Static Charge in 5D Schwarzschild-Tangherlini Spacetime}
We will now derive coordinate expansions for the singular field of a static electric and scalar charge in the 5D Schwarzschild-Tangherlini spacetime. For an electric charge, the singular field is given by Eq.~(\ref{eq:singfieldelec}) where the biscalars therein possess expansions of the form (\ref{eq:vwpoissonexp}) and $\tilde{w}(\bx)$ is given by Eq.~(\ref{eq:wreduced}). Moreover, the metric that induces the Poisson form of the electrostatic wave equation is $\tilde{h}_{\alpha\beta}=f^{-1/2}h_{\alpha\beta}$ where we recall that  $f=1-r_{\textrm{\tiny{H}}}^{2}/r^{2}$ and the Ricci tensor and Ricci scalar on this metric are
\begin{align}
\label{eq:ricciconformalelectric}
\tilde{R}_{\alpha\beta}&=\textrm{diag}\Big\{-\frac{3\,r_{\textrm{\tiny{H}}}^{2}}{2 r^{4}f^{2}}(3\,f+1),\, \frac{r_{\textrm{\tiny{H}}}^{2}}{r^{2}f}(3\,f-1)\Omega_{AB}\Big\},\nonumber\\
\tilde{R}&=-\frac{9}{2}\frac{r_{\textrm{\tiny{H}}}^{4}}{r^{6}\sqrt{f}}
\end{align}
where $\Omega_{AB}$ is the metric on the 3-sphere. Combining Eqs.~(\ref{eq:vwpoissonexp}), (\ref{eq:singfieldelec}), (\ref{eq:wreduced}) and (\ref{eq:ricciconformalelectric}) with a coordinate expansion for $\tilde{\sigma}$ yields a coordinate expansion for the singular field. If we specialize to the situation where the field point $\bx$ and charge location $\bx_{0}$ are separated only in the radial direction, then defining $\Delta r=r-r_{0}$ and expanding the expression (\ref{eq:singfieldelec}) in powers of $\Delta r$ gives
\begin{widetext}
\begin{align}
\label{eq:elecsingfieldexp}
\Phi_{\textrm{\tiny{S}}}(r)&=-\frac{e}{2}\frac{f_{0}^{3/2}}{\Delta r^{2}}+\frac{3 e}{4 r_{\textrm{\tiny{H}}}}\frac{y_{0}^{3/2}f_{0}^{1/2}}{\Delta r}+\frac{e}{32 r_{\textrm{\tiny{H}}}^{2}}\frac{y_{0}^{2}}{f_{0}^{1/2}}\Big(32-(29+3\alpha)y_{0}-3y_{0}\log\Big[\frac{3\,y_{0}^{3}\Delta r^{2}}{16 r_{\textrm{\tiny{H}}}^{2}f_{0}^{2}}\Big]\Big)\nonumber\\
&+\frac{e}{64r_{\textrm{\tiny{H}}}^{3}}\frac{y_{0}^{5/2}\Delta r}{f_{0}^{3/2}}\Big(80-6(26+3\alpha) y_{0}+(71+15\alpha) y_{0}^{2}-3y_{0}(6-5y_{0})\log\Big[\frac{3\,y_{0}^{3}\Delta r^{2}}{16 r_{\textrm{\tiny{H}}}^{2}f_{0}^{2}}\Big]\Big)+\textrm{O}(\Delta r^{2})
\end{align}
\end{widetext}
where $f_{0}=f(r_{0})$ and we have expressed our expansion in terms of an inverse dimensionless radius
\begin{align}
y_{0}=\frac{r_{\textrm{\tiny{H}}}^{2}}{r_{0}^{2}}.
\end{align}

For the static scalar charge in a 5D Schwarzschild-Tangherlini spacetime, the singular field is given by Eq.~(\ref{eq:singfieldscalar}) with $N_{0}=\sqrt{f_{0}}$. The metric that induces the Poisson form of the static scalar wave equation is $\tilde{h}_{\alpha\beta}=f^{1/2}\,h_{\alpha\beta}$ for which the Ricci tensor and Ricci scalar are
\begin{align}
\label{eq:ricciconformalscalar}
\tilde{R}_{\alpha\beta}=\textrm{diag}\Big\{\frac{3\,r_{\textrm{\tiny{H}}}^{4}}{2\,r^{6}\,f^{2}},0,0,0\Big\},\quad\tilde{R}=\frac{3\,r_{\textrm{\tiny{H}}}^{4}}{2\,r^{6}\,f^{3/2}}.
\end{align}
As in the electrostatic case, we combine Eqs.~(\ref{eq:vwpoissonexp}), (\ref{eq:singfieldscalar}), (\ref{eq:wreduced}) and (\ref{eq:ricciconformalscalar}) with a coordinate expansion for $\tilde{\sigma}$ to obtain a coordinate expansion for the singular field, which for radial separation yields
\begin{widetext}
\begin{align}
\label{eq:scalarsingfieldexp}
\varphi_{\textrm{\tiny{S}}}(r)&=\frac{q}{2}\frac{f_{0}}{\Delta r^{2}}-\frac{ q}{4 r_{\textrm{\tiny{H}}}}\frac{y_{0}^{3/2}}{\Delta r}-\frac{q}{96 r_{\textrm{\tiny{H}}}^{2}}\frac{y_{0}^{2}}{f_{0}}\Big(24-(19-3\alpha)y_{0}+3y_{0}\log\Big[\frac{y_{0}^{3}\Delta r^{2}}{16\,r_{\textrm{\tiny{H}}}^{2}f_{0}^{2}}\Big]\Big)\nonumber\\
&-\frac{q}{64r_{\textrm{\tiny{H}}}^{3}}\frac{y_{0}^{5/2}\Delta r}{f_{0}^{2}}\Big(16-2(10-3\alpha)y_{0}+3(3-\alpha)y_{0}^{2}+3y_{0}(2-y_{0})\log\Big[\frac{y_{0}^{3}\Delta r^{2}}{16\,r_{\textrm{\tiny{H}}}^{2}f_{0}^{2}}\Big]\Big)+\textrm{O}(\Delta r^{2}).
\end{align}
\end{widetext}

\begin{figure*}
\centering
\includegraphics[width=11.0cm]{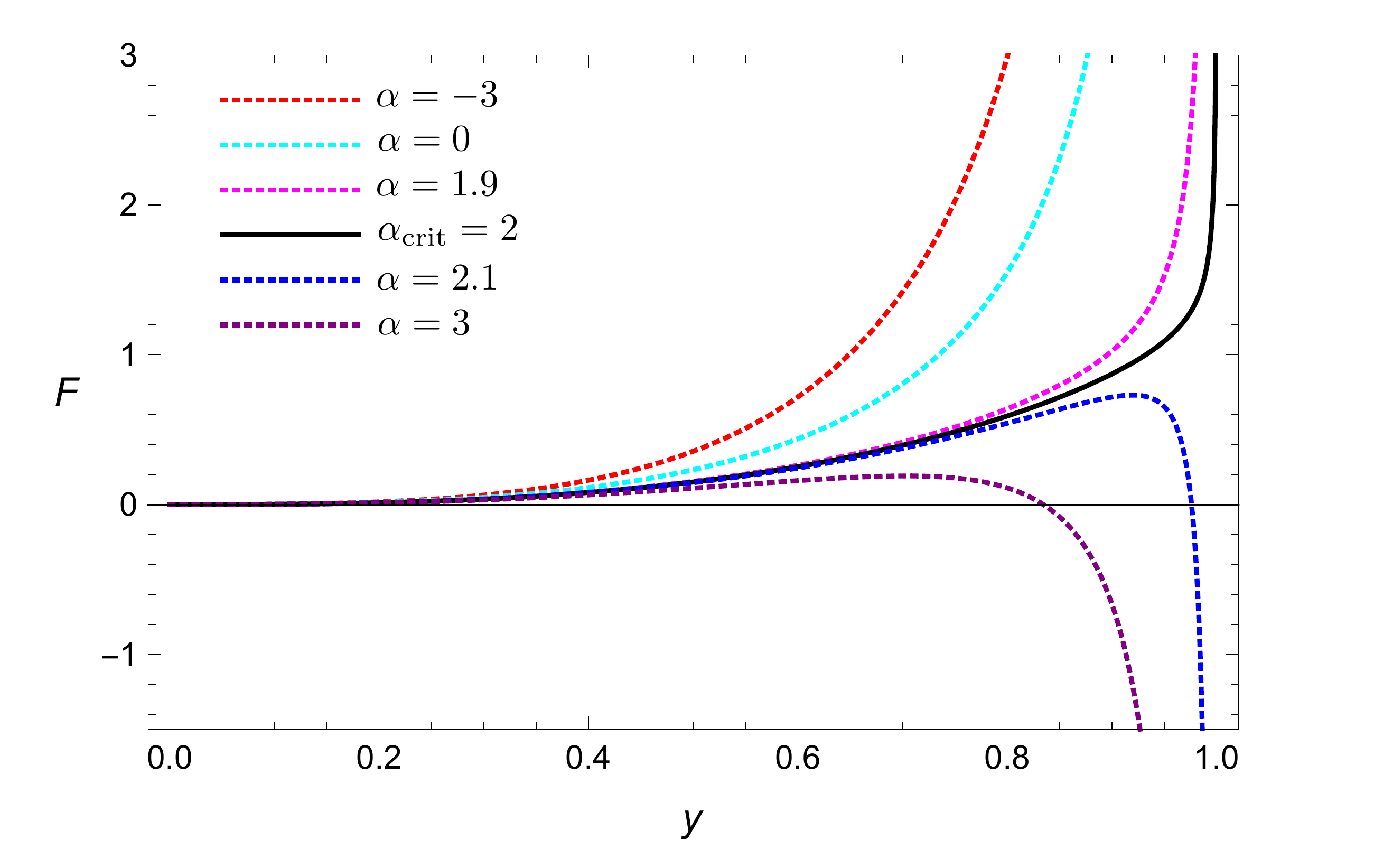}
\caption{{Plot of the self-force invariant on a static electric charge in a 5D black hole spacetime. } We have plotted the force as a function of the inverse dimensionless radius $y=r_{\textrm{\tiny{H}}}^{2}/r^{2}$ for a set of $\alpha$ values. We see that the force diverges for all values of $\alpha$ as the static charge is placed on the horizon $y\to 1$. However, for one particular value, $\alpha_{\textrm{crit}}=2$, the leading order divergence vanishes and we have a milder singularity on the horizon. We also see that the force is repulsive for $\alpha<\alpha_{\textrm{crit}}$, while for $\alpha>\alpha_{\textrm{crit}}$ the force becomes attractive within some radius that depends on $\alpha$. }
\label{fig:selfforce}
\end{figure*} 

\section{Electrostatic Self-Force}
\label{sec:selfforceelec}
In order to compute the self-force, we replace the field gradient in the formal definition (\ref{eq:selfforceelecdef}) with the regularized field gradient before evaluating at the charge's location,
\begin{align}
\label{eq:selfforcereg}
F^{r}=e\,\sqrt{f_{0}}\partial_{r}\Phi_{\textrm{R}}(\bx_{0})=e\,\sqrt{f_{0}}\big[\partial_{r}\Phi-\partial_{r}\Phi_{\textrm{\tiny{S}}}\big]_{r=r_{0}}.
\end{align}
Rather than working with the coordinate-dependent self-force, let us introduce the invariant
\begin{align}
\label{eq:Freg}
F=\pm\sqrt{g_{ab}F^{a}F^{b}}=f_{0}^{-1/2}F^{r}=e\,\big[\partial_{r}\Phi-\partial_{r}\Phi_{\textrm{\tiny{S}}}\big]_{r=r_{0}},
\end{align}
which represents the magnitude of the force measured by an observer at the charge's location. The sign was chosen to coincide with the sign of $F^{r}$. To compute $F$ requires a coordinate expansion for the electrostatic field which we derived in closed form in Sec.~\ref{sec:elecfieldclosed}. Taking the angular coincidence limit $\gamma=0$ in Eq.~(\ref{eq:elecfieldclosed}) and expanding in powers of $\Delta r$ yields
\begin{widetext}
\begin{align}
\label{eq:elecfieldexp}
\Phi(r)&=-\frac{e}{2}\frac{f_{0}^{3/2}}{\Delta r^{2}}+\frac{3\,e}{4 r_{\textrm{\tiny{H}}}}\frac{y_{0}^{3/2}f_{0}^{1/2}}{\Delta r}-\frac{e}{32 r_{\textrm{\tiny{H}}}^{2}}\frac{y_{0}}{f_{0}^{1/2}}\Big(-4-28 y_{0}+35 y_{0}^{2}+\sqrt{f_{0}}(4+6y_{0})\nonumber\\
&+6y_{0}^{2}\log\Big(\frac{y_{0}(1+\sqrt{f_{0}})}{8\sqrt{f_{0}}}\Big)+3y_{0}^{2}\log\frac{\Delta r^{2}}{r_{\textrm{\tiny{H}}}^{2} f_{0}}\Big)+\frac{e}{64 r_{\textrm{\tiny{H}}}^{3}}\frac{y_{0}^{3/2}\Delta r}{f_{0}^{3/2}}\Big(8+60 y_{0}-180 y_{0}^{2}+101y_{0}^{3}\nonumber\\
&-\sqrt{f_{0}}(8+16y_{0}-30y_{0}^{2})-6y_{0}^{2}(6-5y_{0})\log\Big(\frac{y_{0}(1+\sqrt{f_{0}})}{8\sqrt{f_{0}}}\Big)-3y_{0}^{2}(6-5y_{0})\log\frac{\Delta r^{2}}{r_{\textrm{\tiny{H}}}^{2} f_{0}}\Big)+\textrm{O}(\Delta r^{2}),
\end{align}
\end{widetext}
where we have again expressed our series coefficients in terms of $y_{0}=r_{\textrm{\tiny{H}}}^{2}/r_{0}^{2}$. Adopting the expansions (\ref{eq:elecsingfieldexp}) and (\ref{eq:elecfieldexp}) in the definition (\ref{eq:Freg}) results in the closed-form expression for the force invariant (where for convenience, we now drop the zero subscript and refer to the particle's coordinate location as $r$)
\begin{align}
F=\frac{e^{2}}{64 r_{\textrm{\tiny{H}}}^{3}}\frac{y^{3/2}}{f^{3/2}}\Bigg(-8+20 y+6(4-3\alpha-3\log 12) y^{2}\nonumber\\
-15(2-\alpha-\log 12) y^{3}+2\sqrt{f}(4+8 y-15y^{2})\nonumber\\
+6y^{2}(6-5y)\arctanh\sqrt{f}\Bigg),
\end{align}
where $y=r_{\textrm{\tiny{H}}}^{2}/r^{2}$. We can simplify the form of this expression by redefining the arbitrary coefficient
\begin{align}
\alpha\to\alpha-\log(12)
\end{align}
whence the self-force assumes the form
\begin{align}
\label{eq:selfforceelec}
F=\frac{e^{2}}{64 r_{\textrm{\tiny{H}}}^{3}}\frac{y^{3/2}}{f^{3/2}}\Big(-8+20 y+6(4-3\alpha) y^{2}-15(2-\alpha) y^{3}\nonumber\\
+2\sqrt{f}(4+8 y-15y^{2})+6y^{2}(6-5y)\arctanh\sqrt{f}\Big).
\end{align}
Evidently, the self-force depends on the arbitrary coefficient $\alpha$, which was inherited from the arbitrariness of $W_{0}(\bx, \bx_{0})$ in the Hadamard regularization prescription. This coefficient is dimensionless and so there are no unresolved length-scales in our expression for the self-force, contrary to the results of Ref.~\cite{FrolovZelnikovSF}. Moreover, our expression for the self-force does not depend on any simple model for the electric charge and is regular for all $r>r_{\textrm{\tiny{H}}}$, contrary to the calculation of Ref.~\cite{Poisson5D}. However, our calculation does not inform us whether the first order self-force in 5D depends on internal structure, since there are dimensionless parameters that characterize properties of a body, and it is conceivable that our variable $\alpha$ could
depend on such parameters.  For example, in 5D flat spacetime one such dimensionless parameter is
\begin{align}
\frac{1}{e^2} \int d^4 x \,j^t(x)\, \Phi^t(x)\, |x|^2,
\end{align}
which arises from the trace part of the quadrupole coupling of the self-energy density of the body with an external gravitational field. Moreover, such dependence on internal structure for spherically symmetric bodies arises in the second-order electromagnetic self force in four dimensions in flat spacetime \cite{FlanaganMoxon}. A hint that it may arise in the first order self-force in five dimensions is provided by the perspective of effective field theory (see Refs.~\cite{GoldbergerRothstein, GalleyHu, GalleyLeibovichRothstein} for the effective field theory approach to the self-force problem). When one writes down all the possible terms in an action for a charged point particle, one term that is allowed is
\begin{align}
c_0 \,e^2 \int  {\vec a}^2\,d\tau
\end{align}
where ${\vec a}$ is the covariant acceleration, $\tau$ is proper time and $e$ is the charge.  (This term will give rise to higher-derivative equations of motion
but a second order in time equation of motion can be obtained by the usual reduction of order technique).  In four dimensions, the parameter $c_0$ has dimensions of length (in units with $c=1$), and so the operator is an irrelevant operator: its effect becomes negligible for very small bodies. In particular $c_0$ would be expected to be of order the size of the body.  However, in five dimensions, $c_0$ is dimensionless, and so the operator is marginal, and should give a non-vanishing contribution to the equations of motion in the limit of small bodies.  The coefficient $c_0$ may be a universal constant, independent of body structure, or it may depend on the body's internal structure. In an upcoming paper \cite{HarteTaylorFlanagan}, we will apply Harte's formalism \cite{HarteReview, Harte2010, Harte2009, Harte2008} to address whether the self-force on static bodies in arbitrary dimensions depends on internal structure and further elucidate the connection between the ambiguity in the choice of the singular field and renormalizations of the body's multipole moments.

The asymptotic form of $F$ at infinity ($y\to 0$) is
\begin{align}
F=\frac{e^{2} r_{\textrm{\tiny{H}}}^{2}}{2 \,r^{5}}+\textrm{O}(r^{-6}),
\end{align}
while near the horizon ($y=1$) we have
\begin{align}
\label{eq:fconservative}
F=\frac{e^{2}}{128 \sqrt{2} r_{\textrm{\tiny{H}}}^{3/2}}\Big(\frac{2-\alpha}{(r-r_{\textrm{\tiny{H}}})^{3/2}}+\frac{158-63\alpha}{4\,r_{\textrm{\tiny{H}}}\,(r-r_{\textrm{\tiny{H}}})^{1/2}}\nonumber\\
+\frac{128\sqrt{2}}{r_{\textrm{\tiny{H}}}^{3/2}}\Big)+\textrm{O}((r-r_{\textrm{\tiny{H}}})^{1/2}).
\end{align}
It is clear that for any choice of $\alpha$, the force invariant diverges as we approach the horizon which is in agreement with the near-horizon behavior found in Ref.~\cite{Poisson5D}. We note that the leading order pathology can be removed by choosing $\alpha=2$ but the subleading divergence survives. Moreover, the force is everywhere repulsive for $\alpha\le 2$ but for $\alpha>2$, the force remains repulsive far from the black hole but is attractive within some critical radius that depends on $\alpha$ (see Fig.~\ref{fig:selfforce}).

That the self-force diverges as the horizon is approached seems intuitive since a particle requires infinite acceleration to remain static on the horizon. For example, the self-force on a static scalar charge diverges on the ergosphere in the Kerr spacetime \cite{OttewillTaylor3} since it requires infinite acceleration to hold the charge fixed there. On the other hand, this ``intuition'' fails in certain cases, for example, the electrostatic self-force invariant in Schwarzschild spacetime is $F=e^{2}M/r^{3}$ \cite{SmithWill}, which is everywhere regular. Presumably, the regularity of the force on the horizon is a coincidence of the Schwarzschild geometry in four dimensions and there appears to be no reason to expect this to be true in general.

\section{Static Scalar Self-Force}
The scalar self-force is defined by
\begin{align}
\label{eq:selfforceregscalar}
F^{r}=q\,f_{0}\,\partial_{r}\varphi_{\textrm{R}}(\bx_{0})=q\,f_{0}\big[\partial_{r}\varphi-\partial_{r}\varphi_{\textrm{\tiny{S}}}\big]_{r=r_{0}}.
\end{align}
The force invariant $F$ is defined by
\begin{align}
\label{eq:finvariantscalar}
F=\pm\sqrt{g_{ab}F^{a}F^{b}}=f_{0}^{-1/2}F^{r}=q\,\sqrt{f_{0}}\big[\partial_{r}\varphi-\partial_{r}\varphi_{\textrm{\tiny{S}}}\big]_{r=r_{0}}.
\end{align}
The radially-separated coordinate expansion for the scalar field is obtained by expanding the closed-form representation (\ref{eq:scalarfieldclosed}) in $\Delta r$ yielding
\begin{widetext}
\begin{align}
\label{eq:scalarfieldexp}
\varphi(r)&=\frac{q}{2}\frac{f_{0}}{\Delta r^{2}}-\frac{q}{4 r_{\textrm{\tiny{H}}}}\frac{y_{0}^{3/2}}{\Delta r}+\frac{q}{32 r_{\textrm{\tiny{H}}}^{2}}\frac{y_{0}}{f_{0}}\Big(-4-4 y_{0}+3 y_{0}^{2}+\sqrt{f_{0}}(4-2y_{0})\nonumber\\
&-2y_{0}^{2}\log\Big(\frac{y_{0}(1+\sqrt{f_{0}})}{8\sqrt{f_{0}}}\Big)-y_{0}^{2}\log\frac{\Delta r^{2}}{r_{\textrm{\tiny{H}}}^{2} f_{0}}\Big)+\frac{q}{64 r_{\textrm{\tiny{H}}}^{3}}\frac{y_{0}^{3/2}\Delta r}{f_{0}^{2}}\Big(-8-4 y_{0}-4 y_{0}^{2}+y_{0}^{3}\nonumber\\
&+\sqrt{f_{0}}(8-8y_{0}+6y_{0}^{2})+6y_{0}^{2}(y_{0}-2)\log\Big(\frac{y_{0}(1+\sqrt{f_{0}})}{8\sqrt{f_{0}}}\Big)+3y_{0}^{2}(y_{0}-2)\log\frac{\Delta r^{2}}{r_{\textrm{\tiny{H}}}^{2} f_{0}}\Big)+\textrm{O}(\Delta r^{2}).
\end{align}
\end{widetext}
Substituting Eqs.~(\ref{eq:scalarsingfieldexp}) and (\ref{eq:scalarfieldexp}) into the definition (\ref{eq:finvariantscalar}) yields (again dropping the subscript $0$ for typographical convenience)
\begin{align}
F=\frac{q^{2}}{64 r_{\textrm{\tiny{H}}}^{3}}\frac{y^{3/2}}{f^{3/2}}\Bigg(8-12 y+6(4-\alpha-\log4) y^{2}\nonumber\\
-(10-3\alpha-3\log4) y^{3}-\sqrt{f}(8-8 y+6y^{2})\nonumber\\
+6y^{2}(2-y)\arctanh\sqrt{f}\Bigg).
\end{align}
Redefining $\alpha$ by
\begin{align}
\alpha\to\alpha-\log 4,
\end{align}
produces a more simple form,
\begin{align}
\label{eq:selfforcescalar}
F=\frac{q^{2}}{64 r_{\textrm{\tiny{H}}}^{3}}\frac{y^{3/2}}{f^{3/2}}\Big(8-12 y+6(4-\alpha) y^{2}-(10-3\alpha) y^{3}\nonumber\\
-\sqrt{f}(8-8 y+6y^{2})+6y^{2}(2-y)\arctanh\sqrt{f}\Big).
\end{align}
As for the electrostatic self-force, the scalar self-force does not have any undetermined length-scales and need not depend on the internal structure of the charge. 
\begin{figure*}
\centering
\includegraphics[width=11.0cm]{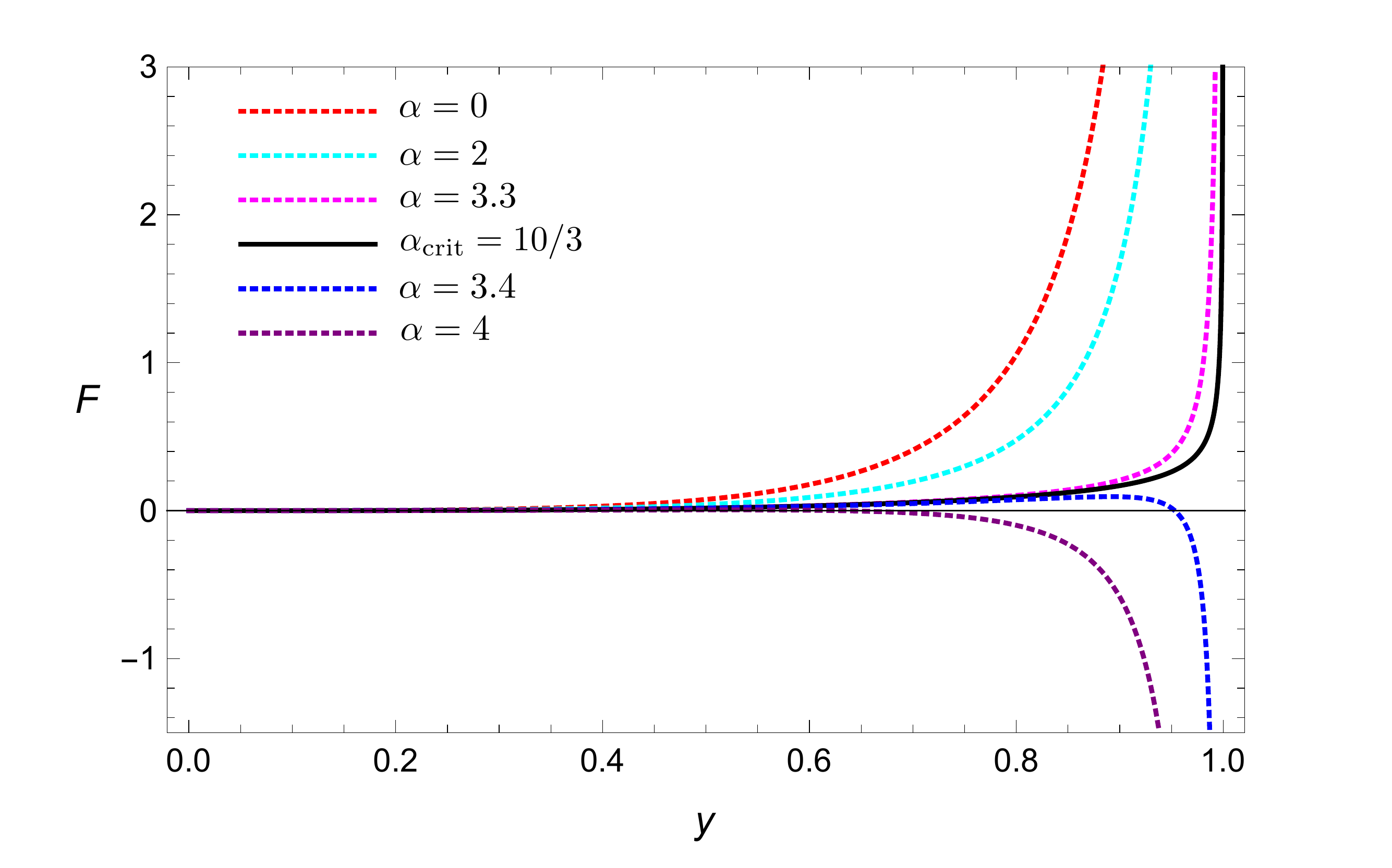}
\caption{{Plot of the self-force invariant on a static scalar charge in a 5D black hole spacetime.} We have plotted the force as a function of the inverse dimensionless radius $y=r_{\textrm{\tiny{H}}}^{2}/r^{2}$ for a series of $\alpha$ values. As in the electrostatic case, the force diverges for all values of $\alpha$ as the static charge is placed on the horizon $y\to 1$, but the divergence is milder for $\alpha_{\textrm{crit}}=10/3$. The force is repulsive for $\alpha<\alpha_{\textrm{crit}}$, while for $\alpha>\alpha_{\textrm{crit}}$, there is a radius which depends on $\alpha$ within which the charge feels an attractive self-force. }
\label{fig:selfforcescalar}
\end{figure*} 

For large $r$, the force invariant behaves as
\begin{align}
F=\frac{3q^{2}r_{\textrm{\tiny{H}}}^{4}}{16 r^{7}}\log(2 r/r_{\textrm{\tiny{H}}})+\textrm{O}(r^{-7}).
\end{align}
Hence, we find the scalar self-force to be repulsive at large $r$ which disagrees with the results of Ref.~\cite{Poisson5D}, though the scaling with $r$ agrees. For the static charge near the horizon, we obtain
\begin{align}
F=\frac{q^{2}}{128\sqrt{2} r_{\textrm{\tiny{H}}}^{3/2}} \Big(\frac{10-3\alpha}{(r-r_{\textrm{\tiny{H}}})^{3/2}}-\frac{78-33\alpha}{4\,r_{\textrm{\tiny{H}}}\,(r-r_{\textrm{\tiny{H}}})^{1/2}}\Big)\nonumber\\
+\textrm{O}((r-r_{\textrm{\tiny{H}}})^{1/2}).
\end{align}
As for the electrostatic case, the force blows up as the static charge is placed closer and closer to the horizon for every choice of the dimensionless coefficient $\alpha$. However, for $\alpha=10/3$ the leading order singularity vanishes and this value delimits the choices for which the force is everywhere repulsive, corresponding to $\alpha\le 10/3$, from those for which the force turns over and becomes attractive as we move towards the horizon, corresponding to parameter values $\alpha>10/3$ (see Fig.~\ref{fig:selfforcescalar}).


\section{Conclusions and Discussion}
In this paper, we have computed closed-form expressions for the self-force on static electric and scalar charges in the 5D Schwarzschild-Tangherlini spacetime. The calculation was facilitated by two results: First, we derived closed-form representations of the electrostatic and scalar static fields. Second, we developed an axiomatic approach to constraining the possible forms of the singular field, reducing the arbitrariness of the Hadamard form to a single arbitrary dimensionless coefficient. From our closed-form expressions for the self-force, Eqs.~(\ref{eq:selfforceelec}) and Eq.~(\ref{eq:selfforcescalar}), we immediately deduce that the self-force does not depend on any unresolved length scales, such as the radius of a sphere centered at the charge's location, nor does our calculation necessitate a dependence of the self-force on the internal structure of the charge. We note that our regularization scheme is quite robust, valid for static charges in arbitrary static 5D spacetimes and easily generalizes to higher dimensional static spacetimes. It relies on three axioms, two of which are quite innocuous assumptions that the singular Green's function scales appropriately under a rescaling of length and is constructed entirely from local geometrical quantities. The third axiom is a strong constraint that rules out ratios of curvature scalars appearing in the singular Green's function, but it would be surprising if the correct expression for the self-force obtained by a matched asymptotics expansion involved such ratios appearing in some special combination that is regular for every static spacetime.

The strong dependence of the self-force on the parity of spacetime dimension is intriguing. It seems intuitive that the singular Green's function should only be sensitive to the local physics in the vicinity of the charge and that it must be symmetric, and yet there is no obvious analog of a Detweiler-Whiting Green's function in odd dimensions satisfying these criteria. Indeed, there are strong indications that such a Green's function does not exist. This raises the very interesting question as to how to regularize the self-force in odd dimensions. Moreover, it remains to be proven whether or not the first order self-force depends on the internal structure of the charge. For static configurations, we address these questions in an upcoming paper \cite{HarteTaylorFlanagan}. In the general dynamical case however, it may be that resolving some of these issues will require a derivation of the self-force using matched asymptotic expansions, or equivalently, by the method of Gralla, Harte and Wald \cite{GrallaHarteWald}.

\section*{Note}
Upon completion of this work, we were made aware of a different derivation by Frolov and Zelnikov of the closed form Green's functions presented here \cite{FrolovZelnikovStaticST1, FrolovZelnikovStaticRN1, FrolovZelnikovStaticRN2}. Their derivation does not rely on summing the modes as we have done but rather relates the static Green's function on the black hole spacetime to the static Green's function on the Bertotti-Robinson spacetime by a simultaneous rescaling of the induced metric and the lapse. Their method is valid for the $d$-dimensional Schwarzschild-Tangherlini or Reissner-N\"ordstrom spacetimes and in general yields a double integral expression for the static Green's function. For $d=5$, the integrals can be written in terms of Elliptic integral functions and we have explicitly checked that our derivation agrees.
\acknowledgements
We are indebted to Eric Poisson, Valeri Frolov, Abraham Harte and Chad Galley for their friendly and insightful correspondence, and to Bernie Nickel for his numerous suggestions for simplifying our closed-form expressions for the static field. We also thank Leo Stein and Justin Vines for many helpful conversations.

P.~Taylor is supported by the Irish Research Council under the ELEVATE scheme which is co-funded by the European Commission under the Marie Curie Actions program. \`E.~\`E. Flanagan acknowledges support from NSF grant PHY-1404105.

\bibliographystyle{apsrev}
\bibliography{database}
\end{document}